\documentclass[%
 aip,
 jcp,
 amsmath,amssymb,
preprint,%
]{revtex4-2}

\usepackage{graphicx}
\usepackage{dcolumn}
\usepackage{bm}
\usepackage[capitalise]{cleveref}
\usepackage[utf8]{inputenc}
\usepackage[T1]{fontenc}
\usepackage{etoolbox}
\usepackage{shortcuts}
\usepackage{verbatim}
\makeatletter
\def\@email#1#2{%
 \endgroup
 \patchcmd{\titleblock@produce}
  {\frontmatter@RRAPformat}
  {\frontmatter@RRAPformat{\produce@RRAP{*#1\href{mailto:#2}{#2}}}\frontmatter@RRAPformat}
  {}{}
}%
\makeatother

\usepackage{amsmath}
\usepackage{amsfonts}
\usepackage{amssymb}
\usepackage{amsbsy}
\usepackage{multirow}
\usepackage{bm}
\usepackage{color}
\usepackage{float}
\usepackage{ulem}
\usepackage{verbatim}

\graphicspath{{./},{./Figures/}}

\begin{document}


\title{
Relating solute interactions to interfacial properties 
}
\author{Varun Mandalaparthy$^*$}
\altaffiliation{Current affiliation: Department of Chemistry, Technical University of Darmstadt, 64287 Darmstadt, Germany}
\affiliation{Department of Chemistry, The Pennsylvania State University, University Park, Pennsylvania 16802, USA}

\author{Benjamin M. Curlee$^*$}%
\altaffiliation{Current affiliation: Department of Chemistry, Princeton University, Princeton, New Jersey 08544, USA}
\altaffiliation{*These authors contributed equally.}
\affiliation{Department of Chemistry, The Pennsylvania State University, University Park, Pennsylvania 16802, USA}%

\author{W.G. Noid}
 \homepage{wgn1@psu.edu}
\affiliation{Department of Chemistry, The Pennsylvania State University, University Park, Pennsylvania 16802, USA}%

\date{\today}

\begin{abstract}
Liquid interfaces are both ubiquitous and also critically important for many commercial products, modern technologies, biological processes, and environmental phenomena. The properties of these interfaces can depend quite sensitively upon the composition of the bulk liquid phase. In this work, we develop a dilute solution theory (DST) for the influence of dilute cosolutes upon the interface between coexisting liquid and vapor phases. We employ a grand canonical perturbation theory to rigorously relate the interfacial free energy to the concentration of the liquid solution. We leverage a corresponding Gibbs ensemble to treat liquid-vapor coexistence and to eliminate the contribution of the bulk phases from this free energy. We express the coefficients of the resulting expansion in terms of microscopic partition functions. By treating solute-solute interactions to lowest order, we distinguish between the intrinsic and effective interfacial preferences of solutes. While the former reflects only solute-solvent interactions, the latter depends upon the solution composition and reflects the influence of solute-solute interactions. We assess this DST with molecular dynamics simulations of binary and ternary mixtures of Lennard-Jones spheres. These simulations demonstrate that DST accurately models the interfacial properties of these systems up to relatively high concentrations. Moreover, the simulations illustrate the impact of attractive solute-solute interactions in converting weak intrinsic surfactants into weak effective depletants.
\end{abstract}

\maketitle

\section{Introduction}

Fluid interfaces are not only of fundamental interest, but also of tremendous practical significance for myriad commercial products,\cite{porter1991, ranji2019, ceresa2021,banat2000} industrial processes,\cite{bhardwaj1993, nelson1982} and environmental phenomena.\cite{donaldson2006, king2009,freedman2017} These interfaces are not merely passive bystanders, but play an essential role in mediating the transfer of energy and matter between coexisting phases.\cite{ maa1983,jeong2012, hsieh2013, stephan2021} Moreover, the asymmetry of fluid interfaces provides a unique environment that can dramatically alter chemical reactivity.\cite{benjamin2015, kumar2018, kusaka2021, ammannSolvationSurfacePropensity2022,martins-costaElectrostaticsChemicalReactivity2023, lacourRoleInterfacesCharge2025}

Accordingly, there exists a vast literature of computational methods and theories for modeling interfacial phenomena.\cite{ gibbs1874, defay1966, adamsonPhysicalChemistrySurfaces,rowlinson2002, taeibirahniComputationalMethodsComplex2015} As recently reviewed,\cite{ kleinheins2023} many studies employ thermodynamic theories to model and interpret experimental measurements.\cite{ butler1932, eberhart1966,shereshefsky1967, connors1989,lamperski1991, chunxi2000, shardt2017, elliott2020, shardt2021, kaptay2018, kleinheins2024} Classical density functional theory provides another computationally efficient framework for modeling macroscopic interfacial properties.\cite{ cahn1958,careyGradientTheoriesFluid1978, evansNatureLiquidvapourInterface1979, wuDensityFunctionalTheory2006, llovellClassicalDensityFunctional2010, rehnerSurfactantModelingUsing2021} Conversely, molecular dynamics (MD) and Monte Carlo (MC) simulations provide a more detailed framework for studying interfaces, albeit at the cost of considerably greater computational expense.\cite{ harrisLiquidvaporInterfacesAlkane1992, sidesCapillaryWavesLiquidvapor1999,ghoufi2008, ghoufi2016, chioMolecularSimulationsFluid2025} Statistical mechanics then provides a framework for bridging molecular interactions to macroscopic interfacial properties.\cite{ kirkwoodStatisticalMechanicalTheory1949, irvingStatisticalMechanicalTheory1950, defay1950, defay1966, englert-chwoles1958, davis1975, dutcher2010,wexler2013}
In particular, a pioneering early study by Guggenheim employed a grand canonical framework and random mixing approximations to model the surface layer above a regular solution.\cite{guggenheim1945}
Subsequently, Ono and Kondo employed a grand canonical perturbation theory to express the interfacial tension as a function of solute activity in terms of cluster integrals.\cite{ono1960}

We have recently developed a lattice-based, dilute solution theory (DST) for modeling the influence of cosolutes upon liquid interfaces.\cite{mandalaparthy2022,mandalaparthy2022a} 
In contrast to many previous studies, we did not adopt any mean field or random mixing approximations. 
Rather we employed a grand canonical perturbation formalism that is somewhat similar to the early work of Ono and Kondo.\cite{ono1960} 
However, in contrast to their work, we explicitly recast the perturbation theory in terms of the bulk solution composition instead of the solute activities. 
Instead of employing cluster integrals, we expressed the coefficients of this perturbation expansion as ratios of partition functions that are amenable to free energy calculations for arbitrarily complex solutes. 
Moreover, by treating solute-solute interactions at lowest order, we derived simple expressions for the surface tension and surface excess. 
These expressions highlight an important distinction between the intrinsic and effective interfacial preference of solutes. 
The intrinsic preference reflects the interactions of infinitely dilute solutes with solvent in bulk and interfacial environments. 
Conversely, the effective preference depends upon the solution composition and reflects solute-solute interactions in the bulk and interfacial environments.

We also reported lattice MC simulations that assessed this DST for binary and ternary solutions.\cite{mandalaparthy2022,mandalaparthy2022a} 
These simulations demonstrated that DST can provide a quantitatively accurate description of the surface tension and surface excess up to surprisingly high concentrations. 
Moreover, these studies clearly demonstrated the influence of cosolute interactions upon effective interfacial preferences. 
In particular, we demonstrated that attractive solute-solute interactions can convert intrinsic surfactants into effective depletants. 
This transition is particularly interesting, as recent studies have highlighted nonadditive effects in cosolute mixtures.
For instance, mixtures of salts exert nonadditive effects on polymer stability\cite{zajforoushanmoghaddam2015, bruce2019, balos2017, bui2021, johnson2019, zhao2022} and on the pH of solutions containing weak acids.\cite{mandalaparthy2026} 
Furthermore, in contrast to earlier suggestions,\cite{lin1994, mello2003, auton2004, holthauzen2006, auton2007} recent studies have reported nonadditive interactions in mixtures of osmolytes,\cite{hunger2015,narang2017, ganguly2020} which are neutral cosolutes that maintain the osmotic balance in marine animals.\cite{yancey1979, yancey2005} 

Here we derive an off-lattice DST for the interfacial properties of inhomogeneous systems. 
We explicitly treat coexistence between two fluid phases and employ the Gibbs ensemble\cite{Panagiotopoulos01071987,panagiotopoulosPhaseEquilibriaSimulation1988} to treat bulk effects from the coexisting phases. 
The resulting expressions for the surface tension and surface excess mirror our previous results for lattice models. 
However, now the key thermodynamic parameters are expressed in terms of ratios of partition functions for an inhomogeneous system and for a Gibbs ensemble. 
We develop simple approximations for evaluating the Gibbs ensemble contributions from simpler 1-phase simulations.  
We validate this theory with MD simulations of binary and ternary mixtures of Lennard-Jones spheres.
These simulations demonstrate that the off-lattice theory quite reasonably describes the conversion of intrinsic surfactants into effective depletants in ternary solutions.

The remainder of this paper is organized as follows. 
Section~\ref{sec:theory} derives the off-lattice DST. 
Section~\ref{sec:methods} summarizes the details of computational studies assessing this theory.
Section~\ref{sec:results_discussion} reports and discusses the results of these computational studies. 
Section~\ref{sec:conclusions} briefly summarizes the conclusions of this paper. 
The supplementary material (SM)  presents a more detailed derivation of the off-lattice theory, as well as additional computational details.

\section{Theory}
\label{sec:theory}

\newcommand{\Dlo}{\De_{\lo}}
\newcommand{\Dlop}{\De_{\lop}}
\newcommand{\Dloop}{\De_{l \oop}}
\newcommand{\Dlnull}{\De_{l \phi}}

\begin{figure}[h]
    \centering
    \includegraphics[width=0.5\linewidth]{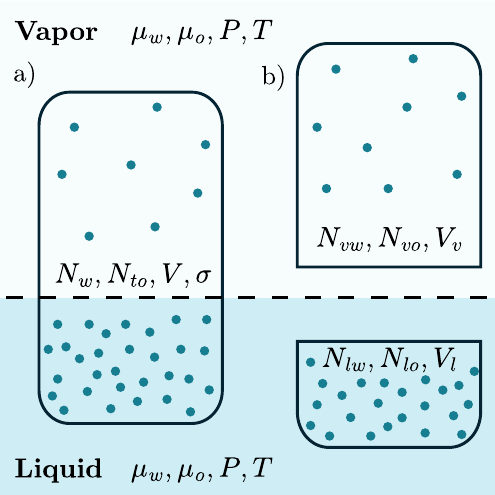}
    \caption{
    Schematic illustrating the total and bulk systems. 
    a) The total system consists of a single region with a fixed volume that includes the liquid-vapor interface.
    b) The bulk system consists of separate liquid and vapor regions that are both far from the liquid-vapor interface.
    }
    \label{fig:theory_fig}
\end{figure}

We develop a simple theory for the influence of cosolute interactions upon the liquid-vapor interface.
We consider solutions with a single dominant solvent, which we label $w$, and $M$ dilute solute species, which we label $o = 1, \ldots, M$.
We employ statistical mechanical perturbation theory to derive the surface tension as a quadratic function of the solution concentration.
The coefficients of this function are expressed in terms of partition functions for 1 or 2 solutes in analogy to the virial series for dilute gases. 
Our approach is inspired by Hill's constant pressure perturbation theory for homogeneous solutions.\cite{hillTheorySolutionsI11957,Hill:1987a}
However, we apply this framework to the inhomogeneous ``total'' system illustrated in Fig.~\ref{fig:theory_fig}a.
Moreover, in order to isolate interfacial properties, we eliminate the contributions from the coexisting liquid and vapor phases. 
We eliminate these contributions by introducing a ``bulk'' system that is illustrated in Fig.~\ref{fig:theory_fig}b and which corresponds to the Gibbs ensemble pioneered by Panagiotopoulos.\cite{Panagiotopoulos01071987,panagiotopoulosPhaseEquilibriaSimulation1988}
We also develop simple approximations for estimating these bulk contributions from single phase simulations. 
We first treat the bulk contributions in Section~\ref{Subsec-Theory-Bulk} and then treat the inhomogeneous total system in Section~\ref{subsec-theory-total}.
While we focus on liquid-vapor interfaces, we anticipate that this framework should readily generalize for any interface between coexisting fluid phases.

\subsection{Bulk system}
\label{Subsec-Theory-Bulk}

\subsubsection{Gibbs ensemble}
\label{SubSec-Theory-GibbsEnsemble}

We define the bulk system by considering the two subsystems in Fig.~\ref{fig:theory_fig}b that are far from the interface.
The vapor subsystem contains $\Nvw$ solvent molecules and $\setNvo$ $o$-solute molecules in a volume, $\Vv$, that is located in the bulk vapor region.
The liquid subsystem contains $\Nlw$ solvent molecules and $\setNlo$ $o$-solute molecules in a volume, $\Vl$, that is located in the bulk liquid region.
We define a closed bulk ensemble by allowing the volumes and compositions of the two subsystems to fluctuate, while holding fixed the total volume, $V$, the total number, $\Nw$, of solvent molecules, and the total number, $\Nbo$, of each $o$-solute species, i.e., 
\begin{equation}
\label{eq-bulk-constr}
\begin{array}{lcl}
\Vv	 + \Vl 	& = & V 	= \text{const}		\\
\Nvw	 + \Nlw 	& = & \Nw = \text{const}		\\
\Nvo + \Nlo 	& = & \Nbo = \text{const} \hspace{0.5in} \text{for $o = 1, \ldots, M$}.
\end{array}
\end{equation}
The partition function for this closed bulk ensemble is 
\begin{eqnarray}
\Qb(\setNbo) 
	& \equiv &  
\label{def-Qb}
		\sum_{\Vl} \sum_{\Nlw}  \sum_{\setNlo} 
			\Ql(\Nlw,\setNlo,\Vl) 
			\Qv(\oNvw,\{\oNvo\},\oVv)	,
\end{eqnarray}
where the sums are restricted $0 \le \Vl \le V$, $0 \le \Nlw \le \Nw$, and $0 \le \Nlo \le \Nbo$ for $o = 1, \ldots, M$.
Here $\Ql$ and $\Qv$ are conventional canonical partition functions for the liquid and vapor subsystems, respectively, while $\oNvw = \Nw - \Nlw$, $\oNvo = \Nbo - \Nlo$, and $\oVv = V - \Vl$ are vapor variables determined by the values of liquid variables and the fixed constraints. 
Here and in the following, we shall implicitly treat $\Nw$, $V$, and $T$ as fixed constants.
Given a fixed number, $\setNbo$, of bulk $o$-solutes, the probability, $\mPb$, that the liquid region contains $\Nlw$ solvent molecules and $\setNlo$ $o$-solute molecules in a volume $\Vl$ is 
\begin{eqnarray}
\label{def-Pb}
\mPb(\Nlw,\setNlo,\Vl; \setNbo) 
	& \equiv & 
		\Ql(\Nlw,\setNlo,\Vl) \times \Qv(\oNvw,\{\oNvo\},\oVv)	 / \Qb(\setNbo)		.
\end{eqnarray}
The SM demonstrates that $\mPb$ is maximized when the two subsystems are in thermodynamic equilibrium. 
Moreover, we expect that when $\Nw$, $V$, and any finite $\Nbo$ are sufficiently large, $\mPb$ will be very sharply peaked about this equilibrium state.
As already noted, this closed bulk ensemble corresponds to the Gibbs ensemble.\cite{Panagiotopoulos01071987,panagiotopoulosPhaseEquilibriaSimulation1988}

\subsubsection{Semigrand bulk ensemble}
\label{SubSec-Theory-SemiGrandBulkEnsemble}
The Helmholtz potential for the bulk system is 
\begin{equation}
\Ab(\setNbo) = - \kT \ln \Qb(\setNbo) = - P V + \muw \Nw + \sumo \muo \Nbo ,
\end{equation}
where the sum runs over the $M$ solute species, $o = 1, \ldots, M$.
In particular, 
$\Abnull = -\kT \ln \Qbnull = -\Pnull V + \muwnull \Nw $ is the Helmholtz potential for a bulk system of pure solvent, while $\Pnull$ and $\muwnull$ are the pressure and solvent chemical potential for liquid-vapor coexistence in the absence of $o$-solutes.
Here we have denoted the canonical partition function for a bulk system with only solvent by 
$\Qbnull = \Qb(\{\Nbo=0\})$.
We define $\Qbo$ as the canonical partition function for a bulk system with a single $o$-solute that may freely partition between the two regions.
Similarly, we define $\Qboop$ as the bulk canonical partition function for exactly two solutes that are of types $o$ and $o'$.
We define  effective partition functions for 1 and 2 solutes in the bulk system:
\begin{eqnarray}
\label{def-qbo}
\qbo 		& \equiv & 	\Nw\inv \Qbo / \Qbnull		\\
\label{def-Zboop}
Z_{b\oop}	& \equiv & 	\Coop \qbo\inv \qbop\inv Q_{b\oop} / \Qbnull		,
\end{eqnarray}
where $\Coop = 1 + \dt_{\oop}$ accounts for indistinguishable solutes.

We now consider the semi-grand ensemble for a bulk system that is open to solutes with fixed $\setmuo$.
The relevant free energy is 
\begin{equation}
\tAb(\setmuo) 
	= \Ab(\setNbo)    - \sumo \muo \Nbo 
	= - P    V + \muw    \Nw	.
\end{equation}
The relevant partition function is 
\begin{eqnarray}
\tQb(\setmuo) 
	& = &  
		e^{-\bt \tAb}
	= 
		\sum_{\setNboZero}			\Qb(\setNbo) \prod_o \lao^\Nbo			\\
 	 & = & 
		\Qbnull + \sumo \Qbo \lao + \half \sumoop \lao \Coop \Qboop \la_\op + \mO(\lao^3),
\end{eqnarray}
where $\lao = e^{\bt \muo}$ is the absolute activity of $o$-solutes.
We  isolate the solute contributions to the bulk free energy by defining
\begin{eqnarray}
\label{def-Psib}
\Psi_b(\setmuo) 
	& \equiv & 
			\Qbnull\inv \tQb(\setmuo) = e^{-\bt \dt \tAb}	\\
	& = & 	
			1 
			+ \Nw \sumo \abo 
			+ \half \sumoop \abo Z_{b \oop} \abop + \mO(\abo^3) 	.
\end{eqnarray}
Here we have introduced the $o$-solute activity for the bulk system,
\begin{equation}
\label{def-abo}
\abo \equiv \qbo \lao		 ,
\end{equation}
and
\begin{equation}
\dt\tAb 
	\equiv 
		\tAb(\setmuo) - \Abnull 
        =
		- V \dt   P   + \Nw\dt\muw ,
\end{equation}
which reflects the influence of $o$-solutes upon the pressure, $\dt P = P - \Pnull$, and the solvent chemical potential, $\dt \muw = \muw - \muwnull$, in the two coexisting phases.
From Eq.~\eqref{def-Psib} we obtain an expansion for the bulk free energy in terms of the bulk solute activity:
\begin{equation}
\label{eq-lnPsib}
- \Nw\inv \bt\dt\tAb 
	= \Nw\inv \ln \Psi_b 
	= \sumo \abo - \half \sumoop \abo \oveboop \abop + \mO(\abo^3) 
    ,
\end{equation}
where 
\begin{equation}
\label{def-oveboop}
\oveboop \equiv - \Nw\inv \left( Z_{b \oop} - \Nw^2 \right) 	
\end{equation}
describes the effective interaction between infinitely dilute $o$- and $o'$-solutes in the bulk system in analogy to the second virial coefficient for dilute gases.\cite{Hill:1987a}

\subsubsection{Approximate treatment of liquid-vapor coexistence}
\label{SubSubSec-TheoryCoexistence}
We now develop a simple approximation for relating the properties of the bulk ensemble to properties of the coexisting liquid and vapor phases.

In the bulk ensemble for pure solvent, the liquid region contains (on average) $\Nstlwz$ solvent molecules in a volume, $\Vl$, that fluctuates about an average $\Vstlz$.
For large systems, we expect that $\Vstlz$ corresponds to the average volume for $\Nstlwz$ solvent molecules at the pressure, $\Pnull$, 
i.e.,  $\Vstlz = \Vstlz(\Nstlwz,\Pnull)$. 
We define an effective partition function, $\qlo$, for an infinitely dilute $o$-solute under these conditions:
$\qlo 
	\equiv 
		\frac{1}{\Nstlwz} \Dlo/\Dlnull 
		$,
where $\Dlnull=\Dnull(\Nstlwz,\Pnull)$ and $\Dlo=\Do(\Nstlwz,\Pnull)$ are the isothermal-isobaric partition functions for pure solvent and for a solution with a single $o$-solute, respectively.
In analogy to Eq.~\eqref{def-oveboop}, we define an effective interaction parameter for infinitely dilute solutes in the liquid phase:
\begin{equation}
\label{def-oveloop}
\oveloop 
	\equiv 
		- \Nstlwz
			\left[
				\Coop \frac{\Dloop \Dlnull}{\Dlo\Dlop} - 1
			\right],
\end{equation}
where $\Dloop = \Doop(\Nstlwz,\Pnull)$.
Following Hill's constant pressure solution theory,\cite{hillTheorySolutionsI11957,Hill:1987a} the activity of an $o$-solute in the liquid phase is 
\begin{equation}
\label{eq-alo}
\alo 
	= 
		\qlo \lao
	= 
		\mo
			\left(
				1 + \sumop \oveloop \mop    
			\right) 	
            + 
            \mO(m^3),
\end{equation}
where $\mo = \Nlo/\Nlw$ is the (dimensionless) molality of $o$-solutes in the liquid phase.
For sufficiently large systems, we expect that $\qlo$ and $\oveloop$ should be independent of system size, i.e.,
$\qlo = \qlo(\Pnull)$ and $\oveloop = \oveloop(\Pnull)$.
In the present work, we shall treat solute-solute interactions in the liquid solution to lowest order and drop terms of $\mO(m^3)$.

In the bulk ensemble for pure solvent, the coexisting vapor region contains (on average) $\Nstvwz$ solvent molecules in a fluctuating volume with an average $\Vstvz$ that corresponds to the equilibrium volume for $\Nstvwz$ solvent molecules at the coexistence pressure, i.e., $\Vstvz = \Vstvz(\Nstvwz,\Pnull)$.
We define an effective partition function, $\qvo$, for a single $o$-solute in solvent vapor under these conditions: 
$\qvo 
	\equiv 
		\frac{1}{\Vstvz} \Qo /\Qnull$, 
where 
$\Qnull = \Qnull(\Nstvwz,\Vstvz)$ 
and 
$\Qo = \Qo(\Nstvwz,\Vstvz)$ 
are the canonical partition functions for pure solvent vapor and for solvent vapor with a single $o$-solute, respectively.
We expect that, for sufficiently large systems, $\qvo = \qvo(\rhostvwz)$, 
where $\rhostvwz = \Nstvwz/\Vstvz = \rhostvwz(\Pnull)$ is the density of pure solvent vapor at the coexistence state point.
In the present work, we shall neglect solute-solute interactions in the vapor phase because we expect they should contribute little for systems far from the critical point.
Consequently, the activity of an $o$-solute in the vapor phase is simply the corresponding density, $\rhovo =  \Nvo / \Vv$, i.e., 
\begin{equation}
\label{eq-avo}
\avo = \qvo \lao = \rhovo   .   
\end{equation}

We now relate the bulk activity, $\abo$, of $o$-solutes to their activity, $\alo$, in the liquid solution:
\begin{equation}
\label{def-alpo}
\alpo \equiv \frac{\qbo}{\qlo} = \frac{\abo}{\alo}		
											.
\end{equation}
The SM demonstrates that 
\begin{equation}
\label{eq-alpo-approx}
\alpo 
	= 	
		\flwnull + \fvwnull \rKo
\end{equation}
where $\flwnull = \Nstlwz/\Nw$ and $\fvwnull = \Nstvwz/\Nw$ are the (average) fraction of solvent molecules in the liquid and vapor phases of the bulk ensemble for pure solvent.
We have also introduced the equilibrium constant
\begin{equation}
\label{def-rKo-maintext}
\rKo 
	\equiv 
		\frac{1}{\rhostvwz} \frac{\qvo}{\qlo} 
		,
\end{equation}
which  depends upon the properties of a single $o$-solute and its interactions with pure solvent in the coexisting liquid and vapor phases.
We can relate $\rKo$ to Henry's law,\cite{Hill:1987a}
\begin{equation}
\label{eq-rKo-maintext}
\rKo
	= 
		\frac{\avo/\rhostvwz}{\alo}		
	= 
	\lim_{\{\la_{\op}\rightarrow0\}}
		\left[ 
		\frac{\rhovo/\rhostvwz}{\mo} 
		\right] 	
		,
\end{equation}
where $\rhovo$ and $\mo$ are the density and molality, respectively, of $o$-solutes in coexisting vapor and liquid phases, while the limit indicates an infinitely dilute solution where $\la_{\op} \rightarrow 0$ for all solute species, $\op$. 
While $\alpo$ reflects the partitioning of the bulk ensemble into liquid and vapor regions, the equilibrium constant $\rKo$ does not.
Equation~\eqref{def-rKo-maintext} also allows us to relate the compositions of the coexisting liquid and vapor phases. 
Given that $\qvo / \qlo = \avo/\alo$, we may use Eqs.~\eqref{eq-alo} and \eqref{eq-avo} to obtain
\begin{equation}
\frac{\rhovo}{\rhostvwz}
    =
        \rKo \mo 
            \left(
                1 + \sumop \oveloop \mop
            \right) 
            .
\end{equation}

The SM derives a simple approximation for the bulk interaction parameter,
\begin{equation}
\label{eq-oveboop-approx}
\oveboop 
	\approx 
			\alpo\inv\alpop\inv \flwnull \oveloop		,
\end{equation}
where $\oveloop$ describes interactions between infinitely dilute solutes in the liquid phase. 
Here we have again neglected the contribution from interactions in the vapor phase.
The factor $\alpo\inv\alpop\inv$ accounts for the tendency of infinitely dilute solutes to partition into the liquid region.
We expect that these arguments can be readily generalized for  any two coexisting fluid phases.

\subsection{Inhomogeneous total system}
\label{subsec-theory-total}
\subsubsection{Semi-grand ensemble for total system}
\label{SubSubSec-Theory-SemiGrandTotal}
We now consider the inhomogeneous total system that is indicated in Fig.~\ref{fig:theory_fig}a.
This total system contains $\Nw$ solvent molecules and $\Nto$ $o$-solutes in a volume, $V$, with an interfacial area, $\si$, at a temperature, $T$.
We shall implicitly treat $\Nw$, $V$, $\si$, and $T$ as fixed constants. 
We  repeat the analysis of Section~\ref{SubSec-Theory-SemiGrandBulkEnsemble} for the inhomogeneous total system.

The Helmholtz potential for the total system is 
\begin{equation}
\At(\setNto) = - \kT \ln \Qt(\setNto) = - P V + \muw \Nw + \sumo \muo \Nto + \ga \si.
\end{equation}
If the total system contains only solvent, then the Helmholtz potential is 
$\Atnull = -\kT \ln \Qtnull = -\Pnull V + \muwnull \Nw + \ganull \si 		,$
where $\ganull$  is the interfacial tension between coexisting solvent phases.
We define corresponding effective one- and two-solute partition functions:
\begin{eqnarray}
\qto 		& \equiv & 	\Nw\inv \Qto / \Qtnull		\\
Z_{t\oop}	& \equiv & 	\Coop \qto\inv \qtop\inv Q_{t\oop} / \Qtnull		,
\end{eqnarray}
where, e.g., $\Qto$ is the canonical partition function when the total system contains a single $o$-solute.

We now consider a semi-grand ensemble that is open to $o$-solutes with chemical potentials, $\setmuo$.
The relevant free energy and partition function are
\begin{eqnarray}
\tAt(\setmuo) 
	& = & \At(\setNto) - \sumo\muo\Nto 
	= -    P V + \muw    \Nw + \ga \si		 \\
\tQt(\setmuo) 
	& = &  
		e^{-\bt \tAt}
	= 
		\sum_{\setNtoZero}	\Qt(\setNto) \prod_o \lao^\Nto			
.
\end{eqnarray}
In order to isolate the contributions from $o$-solutes, we define
\begin{eqnarray}
\label{def-Psit}
\Psi_t(\setmuo) 
	& \equiv & 
			\Qtnull\inv \tQt(\setmuo) = e^{-\bt \dt \tAt}	\\
	& = & 	
			1 
			+ \Nw \sumo \ato 
			+ \half \sumoop \ato Z_{t \oop} \atop + \mO(\ato^3) 	,
\end{eqnarray}
where 
\begin{equation}
\ato \equiv \qto \lao		.
\end{equation}
is the corresponding activity for the inhomogeneous total system
and 
\begin{equation}
\dt\tAt 
	\equiv 
		\tAt(\setmuo) - \Atnull 
	= 
		- V \dt  P    + \Nw\dt\muw    + \si \dt\ga	
\end{equation}
 reflects the influence of $o$-solutes upon the surface tension, $\dt \ga = \ga - \ganull$.
From Eq.~\eqref{def-Psit} we obtain an expansion for the free energy in terms of the solute activity:
\begin{equation}
\label{eq-lnPsit}
-\Nw\inv \bt\dt\tAt 
	= \Nw\inv \ln \Psi_t 
	= \sumo \ato - \half \sumoop \ato \ovetoop \atop + \mO(\ato^3) 
\end{equation}
where 
\begin{equation}
\label{def-ovetoop}
\ovetoop \equiv - \Nw\inv \left( Z_{t \oop} - \Nw^2 \right) 		.
\end{equation}
is the corresponding interaction parameter for infinitely dilute solutes.

\subsubsection{Interfacial properties}
\label{SubSubSec-InterfacialProp}
We now eliminate the bulk contributions to determine the influence of solutes upon the interfacial free energy
\begin{equation}
\dt\tAi 
	\equiv \dt\tAt(\setmuo) - \dt\tAb(\setmuo) 
	= \si \dt\ga 
	= -\kT \ln \left[\Psi_t/\Psi_b\right]		.
\end{equation}
We use the expansions of Eqs.~\eqref{eq-lnPsit} and \eqref{eq-lnPsib} to obtain 
\begin{equation}
-\Nw\inv \bt \dt\tAi 
	= 
		\sumo (\rto - 1) \abo 
	- 
		\half \sumoop \abo \left[ \rto \ovetoop \rtop - \oveboop \right] \abop 	
	+ 
		\mO( \abo^3 ) ,
\end{equation}
where we have defined
\begin{equation}
\label{def-rto}
\rto = \qto / \qbo 
\end{equation}
and used $\ato = \qto \lao = \rto \abo$.
We introduce an area scale $\si_1$ in order to define a dimensionless surface tension, $\bga \equiv \bt \si_1 \ga$, and also a dimensionless parameter, $\eta = \si_1\inv \si / \Nw$, that characterizes the surface-volume ratio.
For instance, if $\si_1$ is the exposed area of a solvent molecule at the liquid-vapor interface, then $\bga$ estimates the contribution of the molecule to the surface free energy (in units of thermal energy) and $\eta$ estimates the fraction of solvent molecules at the interface.
With these definitions, we obtain a simple expression for the scaled surface tension, 
\begin{equation}
\label{eq-dtbga-abo}
-\dt\bga = \sumo \kbo \abo - \half \sumoop \abo \oveioop \abop + \mO(\abo^3) 
\end{equation}
where 
\begin{eqnarray}
\label{def-kbo}
\kbo 
	& \equiv & 
			\eta\inv \left( \rto - 1 \right) 
		= 	
			\eta\inv\left( \frac{\qto - \qbo}{\qbo} \right) 	\\
\label{def-oveioop}            
\oveioop
	& \equiv & 
			\eta\inv\left( \rto \ovetoop \rtop - \oveboop \right) 		.
\end{eqnarray}

\subsubsection{Approximate expressions in terms of liquid composition}
We now employ subsection~\ref{SubSubSec-TheoryCoexistence} to relate interfacial properties to the composition of the liquid solution.
Specifically, Eq.~\eqref{def-alpo} relates $\abo$ to the solute activity in the coexisting liquid phase, $\alo$.
Then Eq.~\eqref{eq-alo} gives $\alo$ as a perturbation expansion in the solute molality in the liquid phase, $\mo = \Nlo / \Nlw$.
This leads to 
\begin{equation}
\label{eq-dtbga-mo}
-\dt\bga 
	= 
		\sumo \ko \mo 
	- 
		\half \sumoop \mo \hoop \mop 
	+ 
		\mO(m^3) 	,
\end{equation}
where
\begin{eqnarray}
\label{def-ko}
\ko 
    & \equiv & 
                \alpo \kbo = \eta\inv \left( \frac{\qto-\qbo}{\qlo} \right)  	\\ 
\label{def-hoop}
\hoop 
	& \equiv & 
        		\alpo \oveioop \alpop - \left( \ko + \kop \right) \oveloop 	.
\end{eqnarray}

Equations~\eqref{eq-dtbga-mo}--\eqref{def-hoop} are a key result of this framework.
They relate the influence of dilute solutes upon the interfacial tension to microscopic partition functions. 
Moreover, Eqs.~\eqref{def-ko} and \eqref{def-hoop} give physically appealing expressions for the two key thermodynamic parameters, $\ko$ and $\hoop$.
The parameter $\ko$ describes the impact of infinitely dilute cosolutes upon the surface tension.
Equation~\eqref{def-ko} expresses $\ko$ as the ratio of the effective one-solute partition function for the interface, $\qio \equiv \qto - \qbo$, and the corresponding effective partition function for the coexisting liquid phase, $\qlo$.
The parameter $\hoop$ describes the impact of cosolute interactions upon the surface tension.
The first term in Eq.~\eqref{def-hoop} scales the interfacial interaction energy, $\oveioop$, by the tendency of the solutes to partition into the liquid phase, while the second term accounts for the bulk interactions that are lost when solutes partition to the interface.
These results are analogous to our earlier results for a simple lattice model that did not treat two-phase coexistence.\cite{mandalaparthy2022,mandalaparthy2022a}

We now seek a simple expression for the surface excess of $o$-solutes.
The Gibbs adsorption equation may be expressed (at constant temperature):
\begin{equation}
\label{eq-scaled-ddtbga}
\dd\dt\bga 
	=
	- 
		\sumo \bGao \dd\bmuo 	,
\end{equation}
where we have scaled the surface excess, $\bGao \equiv \si_1 \Gao = \frac{\si_1}{\si} \left( \Nto - \Nbo \right)$, and solute chemical potential, $\bmuo = \bt \muo$. 
From Eq.~\eqref{eq-dtbga-mo}, we obtain
\begin{equation}
\label{def-bo}
\rb_o 
	\equiv 
			\left( 
				\frac{\ptl\bga}{\ptl\mo}
			\right)_{T,\notmo}
	= 
		-
			\ko 
		+ 
			\sumop 
					\hoop \mop
		+ 
			\mO(m^2) 	,
\end{equation}
where the subscript $\notmo$ indicates that the derivative is evaluated while holding $m_{\op}$ constant for all $\op \neq o$.
For simplicity, we approximate the chemical potential, $\bmuo(T,\setmo)$, of the coexisting liquid phase by the chemical potential, $\bmuo(T,\Pnull,\setmo)$, of the bulk liquid phase at the pressure, $\Pnull$, that corresponds to coexistence for pure solvent.

We then use Eq.~\eqref{eq-alo} to evaluate the derivative
\begin{equation}
\label{def-mKoop}
\mKoop 
	\equiv
		\left( 
			\frac{\ptl \bmuo}{\ptl\mop}
		\right)_{T,\Pnull,\notmop} 
	= 
		m\inv_o \de_{\oop}
		+
		\oveloop  
		+ 
		\mO(m) 
	= 
		\mK_{\op o}	.
\end{equation}
Using this approximation, Eqs.~\eqref{eq-scaled-ddtbga}-\eqref{def-mKoop} imply  
\begin{equation}
\label{eq-bo}
\rb_o 
	\approx
		- 
		\sumop \mKoop	 \bGa_\op		,
\end{equation}
for variations at constant temperature.
We invert this system of equations to determine the surface excess:
\begin{equation}
\bGa 
	\approx
		- \mK\inv \rb		.
\end{equation}
For systems with a single type of $o$-solute, we obtain 
\begin{equation}
\label{eq-bGa-onesolute}
\bGa 
	\approx
		m
		\left( 
			\frac{k-hm}{1+\ove m}
		\right)
		,
\end{equation}
where we have suppressed the subscript $o$.
For ternary solutions with two types of $o$-solutes, $o = 1$ and 2, we  obtain
\begin{equation}
\label{eq-bGa-twosolutes}
\bGa_1 
	\approx
		m_1
		\left[ 
			\frac{
				k_1 
				- h_{11} m_1 
				- \left( h_{12} - k_1 \ove_{22} + k_2 \ove_{12} \right) m_2
				}
				{
				1 + \ove_{11} m_1 + \ove_{22} m_2 
				}
		\right]
\end{equation}
and an analogous equation for $\bGa_2$.

Equations~\eqref{eq-bGa-onesolute} and \eqref{eq-bGa-twosolutes} provide a second key result by explicitly demonstrating the impact of cosolute interactions upon the interfacial composition. 
Moreover, Eqs.~\eqref{eq-bGa-onesolute} and \eqref{eq-bGa-twosolutes} predict these effects in terms of microscopic partition functions.
It is convenient to introduce the ``effective interfacial preference'' for $o$-solutes:
\begin{equation}
\label{def-tko}
\tko 
	\equiv
		 \left( \frac{\ptl \bGao}{\ptl \mo} \right)_{m_{\hat{o}'}}	
		 ,
\end{equation}
which describes the impact of the solution composition upon the fraction of added $o$ solutes that migrate to the surface.
For ternary solutions  
\begin{equation}
\tk_1 
	\approx 
			k_1 
		- 
			2 \left(  h_{11} + k_1 \ove_{11} \right) m_1
		-       
			\left( h_{12} + k_2 \ove_{12} \right) m_2 	
        +
            \mO(m^2)
            \xrightarrow{m_1,m_2\rightarrow0}
    k_1
\end{equation}
and an analogous equation describes $\tk_2$.
Thus, $\ko$ describes the ``intrinsic interfacial preference'' of $o$-solutes of infinitely dilute solutes, while $\tko$ describes how cosolute interactions modulate this intrinsic preference.
Similarly, the theory predicts that the addition of 1-solutes can drive 2-solutes either towards or away from the interface according to
\begin{equation}
\tk'_{2|1}
	\equiv 
		\left( \frac{\ptl \bGa_2}{\ptl m_1} \right)_{m_2}
	\approx 
		- \left( h_{12} + k_1 \ove_{12} \right) m_2 
        +
            \mO(m^2)
            .
\end{equation}

\section{Computational Methods}
\label{sec:methods}

\subsection{Model systems}
We simulated dilute solutions with a dominant solvent species ($w$) and one or two distinct solute species ($o = 1$ or 2). 
We modeled all molecules as spherical particles and described all interactions with simple Lennard-Jones (LJ) potentials\cite{jones1924, jones1924a, schwerdtfeger2024} of the form
\begin{equation}\label{eq:LJ_potential}
    U_2(r) = 4\eps 
    \left[ 
     \left( \frac{\sigma}{r} \right)^{12} 
    -\left( \frac{\sigma}{r} \right)^6 
    \right]    ,
\end{equation}
where $\sigma$ and $\eps$ are the LJ diameter and well depth respectively. 
Note the important distinction between the LJ parameter, $\eps$, and the DST thermodynamic interaction parameters, $\ve$, defined by, e.g., Eq.~\eqref{def-oveloop}. 
For simplicity, we adopted the same LJ diameter for all interactions.
We distinguished different species by the corresponding LJ well depths.

In the following, we report simulation parameters in physical units.
We assumed that all molecules had the same mass, m $ = 1$~amu, and defined the characteristic length scale by setting $\sigma = 1$~nm.
We defined the characteristic energy scale by setting the well-depth of the solvent-solvent potential:  $\eps_{ww} = 1$~kJ/mol.
This determines the corresponding time scale $t = $ 1~ps. 
Corresponding LJ values are obtained by dividing energies by  $\eps_{ww}$ = 1~kJ/mol, lengths by $\sigma$ = 1~nm, and times by $t = $ 1~ps. 
In the case of nontrivial conversions, we report the corresponding LJ parameters in parentheses.

We simulated binary solutions with two classes of solutes that were distinguished by their interaction with solvent. 
Surface active (SA) solutes were characterized by $\eps_{ow} \neq \eps_{ww}$, while surface neutral (SN) solutes were characterized by $\eps_{ow} = \eps_{ww}$.
We considered 4 different SA solutes with $\eps_{ow}/ \eps_{ww} = $ 0.90, 0.95, 1.05, and 1.55.
For simplicity we set $\eps_{oo} = \eps_{ww}$ for each SA solute.
Conversely, we considered 5 different SN solutes with $\eps_{oo}/\eps_{ww} = $ 0.10, 0.50, 0.80, 1.20, and 1.50.
We also simulated a ternary solution with two symmetric cosolute species,  $o = $ 1 and 2.
In this case we set $\eps_{1w} = \eps_{2w} = 0.99 \eps_{ww}$, $\eps_{11} = \eps_{22} = 1.00 \eps_{ww}$, and $\eps_{12} = 1.50 \eps_{ww}$.

\subsection{Simulation details}
\label{subsec-simdetails}

\subsubsection{Simulation parameters}
\label{subsubsec-simparam}
We performed all simulations with GROMACS version 2019.6.\cite{abraham2015}
All simulations employed periodic boundary conditions in all three dimensions.
We propagated dynamics with the velocity Verlet integrator, while employing a 2 fs 
time step and calculating kinetic energies by averaging over half time steps.\cite{Frenkel:2002}
We employed the Verlet cutoff scheme with a 2.5~nm 
cutoff for the neighbor list and  updated this list every time step.
We employed the GROMACS force-switch option to switch forces to zero over the distance range from 2.1~nm 
to 2.5~nm. 
In order to ensure consistency between slab and bulk simulations, we did not include dispersion corrections for the pressure or energy. 
We employed the Bussi thermostat\cite{Bussi2007} to sample canonical energy fluctuations at a temperature $T = $ 65 K 
($\tilde T = $ 0.54) with a time constant of 0.5 ps.
Production simulations in the constant NPT ensemble employed the Martyna–Tuckerman–Tobias–Klein (MTTK) barostat\cite{martyna1996} to maintain a constant external pressure $P_{\rm ext} = $ 0.000427~bar ($\tilde P_{\rm ext} = 2.57\times 10^{-5}$), which corresponds to the pure solvent vapor pressure, $\Pnull$, while employing a time constant of 2.0 ps 
and a compressibility of $4.5\times10^{-5}$ bar$^{-1}$ ($\tilde \kappa_{\rm T} = $ 0.000747).

\subsubsection{Inhomogeneous slab simulations}
\paragraph{Pure solvent slab}
We created a pure solvent slab by first equilibrating a cubic box with 2000 solvent particles for 5 ns 
in the constant NPT ensemble at an external pressure  $P_{\rm ext} = $ 1 bar ($\tilde P_{\rm ext} = $ 0.0602) with the Berendsen barostat.\cite{berendsen1984}
The resulting configuration had dimensions $L_x = L_y = L_z = $ 13.47376~nm.
We employed the GROMACS editconf function to extend the z-dimension of the simulation box by a factor of 4.\cite{Muller:2021wn}
We equilibrated the resulting slab configuration in the constant NVT ensemble for 20 ns.

In order to precisely characterize the pure solvent system, we made 508 copies of the final equilibrated slab configuration.
We simulated each of these 508 replicates in the constant NVT ensemble for 220~ns. 
We discarded the first 40~ns of each replicate simulation and collected statistics from the remaining 180~ns, resulting in a total of 91.44 $\mu$s of production simulation. 
Given these equilibrium simulations, we characterized liquid-vapor coexistence for pure solvent according to the protocols described below. 
We determined the surface tension of pure solvent to be, $\ganull = $ 13.15~bar nm ($\tilde \ga_\phi = $ 1.465).
We determined the coexistence pressure, $\Pnull = $ 0.000427~bar ($\tilde P_\phi = 2.57\times 10^{-5}$), by averaging the $P_{zz}$ component of the pressure tensor.  
We determined the solvent number densities for the coexisting liquid and vapor phases to be 
$\rhostlwz = $ 0.853 nm$^{-3}$ 
and $\rhostvwz = $ 0.00112 nm$^{-3}$, 
respectively. 
From Eq.~\eqref{eq-dividing-surface} below we determined the width of the liquid phase to be $d = $ 12.9~nm.
From this we determined the volumes of the coexisting liquid and vapor phases to be $\Vstlz = L_x L_y d = $ 2330~nm$^3$ and  $\Vstvz = L_x L_y (L_z - d) = $ 7390~nm$^3$, respectively.
This then determined the equilibrium number of solvent molecules in the liquid, $\Nstlwz = \Vstlz\rhostlwz = $ 1992, and vapor, $\Nstvwz = \Vstvz\rhostvwz = $ 8, phases, as well as the corresponding fraction of molecules in the two phases, $\flwnull = \Nstlwz/\Nw = $ 0.996 and $\fvwnull = \Nstvwz/\Nw = $ 0.004.

\paragraph{Slabs of dilute solutions}
We created slabs of liquid solutions by employing the GROMACS `insert-molecules' function to insert an appropriate number of solute molecules into the equilibrated slab of pure solvent.
We simulated 10 replicates for each binary solution with SA solutes, while simulating 64 replicates for each binary solution with SN solutes and for each ternary solution.
We simulated each replicate in the constant NVT ensemble for 190~ns and discarded the initial 40~ns as an equilibration period.
We sampled energies and pressures at every time step from the remaining 150~ns,  
while sampling configurations every 4~ps. 
This provides a total of 1.5 $\mu$s production simulation 
for each binary SN solution and 9.6 $\mu$s production simulation 
for each binary SA solution and for each ternary solution.

\paragraph{Characterizing the coexisting phases}
We employed the GROMACS gmx density program to compute the density profile from each slab simulation.
In particular, we employed the -center option to shift the mass center of each sampled configuration to the origin. 
We then determined the equilibrium densities of the liquid and vapor phases  based upon the central 10~\% and the outer 40~\%, respectively, of the simulated box.
For instance, we determined the linear density, $\la_{li}$, of each component ($i = w$ or $o$) in the liquid phase by counting molecules with z-coordinates in the interval $- 0.05 L_z \le z \le + 0.05 L_z$.
We adopted a Gibbs dividing surface that is equimolar with respect to solvent such that
\begin{eqnarray}
\label{eq-Nw-GibbsDivSurf}
    N_w = \lambda_{vw}(L_z-d)+\lambda_{lw}d ,
\end{eqnarray}
where $d$ is the width of the liquid phase. 
This allows us to determine 
\begin{eqnarray}
\label{eq-dividing-surface}
d=\frac{N_w-\lambda_{vw}L_z}{\lambda_{lw}-\lambda_{vw}} .
\end{eqnarray}
The surface excess of $o$-solute is defined 
\begin{equation}
\label{def-Gao-methods}
\Gamma_o 
    = 
        \si\inv \left[ N_{to}-\overline N_{vo}-\overline N_{lo}\right]       
\end{equation}
where $N_{to}$ is the total number of $o$-solutes in the slab simulation and $\si = 2 L_x L_y$ is the surface area, while $\overline N_{lo} = \la_{lo} d$ and $\overline N_{vo} = \la_{vo} (L_z - d)$ are the average number of $o$ solutes in the liquid and vapor phases, respectively. 
By using Eq.~\eqref{eq-dividing-surface}, we determined the surface excess
\begin{eqnarray}
\Gamma_o 
    = 
        \si\inv 
    \left[N_{to}-\lambda_{vo}L_z-(N_w-\lambda_{vw}L_z)
    \left(\frac{\lambda_{lo}-\lambda_{vo}}{\lambda_{lw}-\lambda_{vw}}\right)\right]
    .
\end{eqnarray}
We determined the molality, $m_o$, of the liquid phase and the surface excess, $\Gao$, for each system by computing these quantities for each independent replicate simulation and then  averaging over the replicates.
We estimated the uncertainty in these quantities by the corresponding standard errors determined from the independent replicate simulations.

\paragraph{Surface tension}
We employed the gmx energy function to calculate the surface tension according to 
\begin{equation}
    \label{def-surf-tens-P}
    \gamma = \frac{1}{2} L_z \left[ P_{zz} - \frac{P_{xx} + P_{yy}}{2} \right]  ,
\end{equation}
where $P_{ii}$ denotes the diagonal element of the pressure tensor in the $i^{\rm th}$ direction.
We employed block averaging to estimate the uncertainty in the calculated surface tension.\cite{flyvbjerg1989}
Specifically, we determined one sample of the surface tension by averaging the calculated surface tension over 2~ps intervals. 
We combined 128 such averages into a single block and determined the block average for the 128 samples.
We determined the mean, $\mu_R$, and variance, $\si^2_R$, across the block averages for each replicate, $R$. 
We estimated the surface tension by averaging the replicate means $\gamma = N_R\inv \sum_{R=1}^{N_R} \mu_R$.
We estimated the uncertainty in the surface tension $\sigma_\ga = \sqrt{N_R^{-2} \sum_{R=1}^{N_R} \si^2_R}$.

\subsubsection{Single phase simulations}
In the present work we determined bulk thermodynamic parameters based upon free energy calculations for separate liquid and vapor phases. 
Specifically, we simulated a bulk liquid of 1994 solvent molecules, which corresponds to $\Nstlwz + 2$.
We simulated this liquid in the constant NPT ensemble at an external pressure corresponding to the vapor pressure of pure solvent, $P_{\rm ext} = \Pnull = $ 0.000427 bar ($\tilde P_{\rm ext} = \tilde P_\phi = 2.57\times 10^{-5}$).
We equilibrated this bulk liquid for 40~ns before performing a 10~ns production simulation, from which we sampled the configuration after every 4~ps.
The average volume of this NPT simulation was $\overline V = \overline L^3$, where $\overline L = $ 13.27~nm. 
This quantitatively reproduced the liquid density of the pure solvent slab simulation, $\rholw = \rhostlwz = $ 0.853~nm$^{-3}$.
Similarly, we performed a constant NVT simulation of 100 solvent molecules at a density corresponding to the vapor density of the pure solvent slab simulation, $\rhovw = \rhostvwz = $ 0.00112 nm$^{-3}$.  
We equilibrated this bulk vapor for 40~ns before performing a 100~ns production simulation, from which we sampled the configuration after every 4~ps.
In cases that the free energy calculations converged sufficiently rapidly, we did not employ all of the sampled configurations.

\subsection{Free energy perturbation}
\label{subsec-FreeEnergyMethods}
DST describes liquid-vapor coexistence and interfacial properties in terms of effective partition functions for 1 or 2 solutes.
In this subsection, we outline the free energy calculations that were employed to determine these effective partition functions.
The SM provides a much more detailed discussion of these calculations.

\newcommand{\Pext}{P_{\rm ext}}
\newcommand{\bV}{\overline V}
\newcommand{\DG}{\De G}
\newcommand{\DGti}{\DG_{\rm TI}}
\newcommand{\muoid}{\muo^{\rm id}}
\newcommand{\muoxs}{\de\muo^{\rm xs}}

\newcommand{\Lao}{\La_o}
\newcommand{\Law}{\La_w}
\newcommand{\hZ}{\widehat Z}

\newcommand{\Yo}{Y_{o}}
\newcommand{\Yoop}{Y_{\oop}}

\subsubsection{One solute effective partition functions}
DST defines several key thermodynamic parameters in terms of effective partition functions, $\qo$, that describe the insertion of a single $o$-solute into various environments.
In particular, $\rKo$ is defined by Eq.~\eqref{def-rKo-maintext} and $\ko$ is defined by Eq.~\eqref{def-ko}.
These effective partition functions also enter the definition of $\alpo$ in Eq.~\eqref{def-alpo} and $\rto$ in Eq.~\eqref{def-rto}.
Each effective one-solute partition function, $\qo$,  is defined by an expression of the form
\begin{equation}
\label{def-qo-methods}
\qo    
    \equiv 
            \frac{1}{\Nw} \frac{\Qo(\Nw)}{\Qnull(\Nw)}
    =
            \frac{1}{\Nw} 
                \frac{\Qo(\Nw)}{\Qnull(\Nw+1)}
                    e^{-\bt \muwnull}   
            ,
\end{equation}
where $\Nw$ is the number of solvent molecules.
We obtain the last expression by introducing $1 = \Qnull(\Nw+1)/\Qnull(\Nw+1)$ and recognizing $\Qnull(\Nw+1,) / \Qnull(\Nw) = e^{-\bt \muwnull}$, where $\muwnull$ is the chemical potential of pure solvent at liquid-vapor coexistence.
Because the solvent chemical potential is uniform throughout the simulated systems, all factors of $e^{-\bt \muwnull}$ cancel from the final expressions for all observables.
The remaining ratio of partition functions may be re-expressed
\begin{equation}
\label{eq-Qo-ratio-methods}
\frac{\Qo(\Nw)}{\Qnull(\Nw+1,)}
    =
        \left(\Nw+1\right) 
            \left( \frac{\Law}{\Lao}\right)^3 
                \Yo(\Nw+1)    ,
\end{equation}
where $\Law$ and $\Lao$ are the thermal de Broglie wavelengths for solvent and $o$-solute molecules, respectively, while $Y_{o}(N+1)$ is defined by the ratio of configuration integrals, $\hZ_{o}(N)/\hZ_{\phi}(N+1)$, associated with converting one solvent molecule into an $o$-solute for a system with $N+1$ solvent molecules:
\begin{eqnarray}
\label{def-Yo}
\Yo(N+1)   
    & \equiv &  
        \frac{\hZ_{o}(N)}{\hZ_{\phi}(N+1)}
    =
      \llangle 
        \frac{1}{N+1}
        \sum_{i=1}^{N+1}
            \exp\left[
                    -\bt \De U_{o|i}(\br^{N+1})
                    \right]
      \rrangle_{\phi, N+1}       .
\end{eqnarray}
The angular brackets in Eq.~\eqref{def-Yo} indicate an average over a system of $N+1$ solvent molecules, while $\De U_{o|i}(\br^{N+1})$ indicates the change in the interaction potential when converting solvent molecule $i$ into an $o$-solute.
This last average may be computed via the standard free energy perturbation (FEP) method, i.e., simulating a system of $N+1$ solvent molecules and averaging   
               $ \frac{1}{N+1}
                    \sum_{i=1}^{N+1}
                    \exp\left[
                    -\bt \De U_{o|i}(\br_t^{N+1})
                    \right]$
over the sampled configurations, $\br_t^{N+1}$.
One expects that, for sufficiently large systems, this ratio should depend only upon the intensive state.

\subsubsection{Two solute effective partition function}
DST defines the solute-solute interaction parameters, $\ove_{\oop}$, and, thus, the interfacial parameter, $h_{\oop}$, in terms of effective partition functions, $Z_{\oop}$, describing the insertion of two cosolutes into various pure solvent environments. 
In each case, the effective two solute partition function is of the form 
\begin{eqnarray}
Z_{\oop}	
    & \equiv & 	
        \Coop \qo\inv \qop \inv 
            \frac{Q_{\oop}(\Nw)}{\Qnull(\Nw)}
    =
        \Coop \qo\inv \qop\inv 
            \frac{Q_{\oop}(\Nw)}{\Qnull(\Nw+2)}
            e^{-2\bt\muwnull}
        ,
\end{eqnarray}
where we have introduced $1 = \Qnull(\Nw+2)/\Qnull(\Nw+2)$ and recognized $\Qnull(\Nw+2) / \Qnull(\Nw) = e^{-2 \bt \muwnull}$.
We express the remaining ratio of partition functions in analogy to Eq.~\eqref{eq-Qo-ratio-methods}:
\begin{eqnarray}
\Coop \frac{Q_{\oop}(\Nw)}{\Qnull(\Nw+2)}
    =
         \left(\frac{(\Nw+2)!}{\Nw!}\right) 
            \left( \frac{\Law^2}{\Lao\La_{o'}}\right)^3 
                \Yoop(\Nw+2)    ,
\end{eqnarray}
where 
\begin{eqnarray}
\Yoop(N+2)    
    \equiv
        \frac{\hZ_{oo'}(N)}{\hZ_{\phi}(N+2)}
    =
      \llangle 
        \frac{N!}{(N+2)!}
        \sum_{i=1}^{N+2}
        \sum_{j (\neq i)}^{N+2}
            \exp\left[
                    -\bt \De U_{oo'|ij}(\br^{N+2})
                    \right]
      \rrangle_{\phi, N+2}       
      . 
      \end{eqnarray}
Here $\De U_{oo'|ij}(\br^{N+2})$ is the change in the interaction potential when solvent molecules $i$ and $j$ are  converted to $o$ and $o'$-solutes, respectively. 
The two-solute effective partition function may then be expressed
\begin{eqnarray}
\label{eq-Zoop}
Z_{\oop} 
    = 
        \Nw^2
        \left( \frac{\Nw+2}{\Nw+1} \right)
        \frac{\Yoop(\Nw+2)}
                {\Yo(\Nw+1)Y_{o'}(\Nw+1)}    
        .
\end{eqnarray}
We found that Eq.~\eqref{eq-Zoop} converged much more rapidly when we calculated $\Yo$ and $\Yoop$ from the same set of configurations. 
Accordingly, we directly calculated $\Yo(\Nw+2)$ and estimated $\Yo(\Nw+1)$ via extrapolation.
The SM describes this extrapolation in detail.

\subsection{Thermodynamic integration for $\muoxs$}
\label{SubSec-Methods-TI}
We employed thermodynamic integration (TI) to estimate the solute chemical potential.
Specifically, we considered a system containing $\Nw = $ 1000 solvent molecules and $\No + 1$ $o$-solutes at a constant external pressure $\Pext = \Pnull$, corresponding to the coexistence pressure for pure solvent.
In the initial equilibrium state ($\la = 0$) one of the $o$-solutes did not interact with the rest of the system, i.e., the system consisted of a solution with $\Nw$ solvent molecules, $\No$ $o$-solutes, and a single ideal gas molecule.
The final equilibrium state ($\la = 1$) corresponded to a fully interacting solution of $\Nw$ solvent molecules and $\No+1$ $o$-solutes.
The change in the Gibbs potential for this transformation may be approximated
\begin{equation}
\DG(\No) \approx \muo(\mo) - A_{\rm IG}(1,\bV_0)
\end{equation}
where $\mo = \No/\Nw$, $A_{\rm IG}(1,\bV_0) = - \kT \ln\left( \bV_0/\La_o^3\right)$ is the Helmholtz free energy for the uncoupled $o$-solute, $\La_o$ is the thermal de Broglie wavelength for the $o$-solute, and $\bV_0$ is the average volume of the system in the initial equilibrium state.  
This Gibbs free energy change can be estimated via TI according to
\begin{equation}
\DG(\No) = \DGti(\No) + \kT \ln \left( \No + 1 \right), 
\end{equation}
where 
\begin{equation}
\DGti(\No) 
    \equiv 
        \int_0^1 \dd \la
                \llangle 
                    \frac{\ptl U(\la)}{\ptl \la} 
                \rrangle_{\Nw,\No+1,\la,\Pnull}    
\end{equation}
and the subscripted angular brackets denote an average over the isothermal-isobaric ensemble for $\Nw$ solvent molecules and $\No+1$ $o$-solutes  with the interaction potential $U(\la)$.
We partitioned the $o$-solute chemical potential $\muo = \muoid + \muoxs$, where $\muoid \equiv \muost + \kT \ln \mo$ is the ideal contribution and $\muoxs$ is the excess contribution due to solute-solute interactions. 
The excess contribution can then be estimated
\begin{equation}
\muoxs(\mo) \approx \DGti(\No) - \DGti(0)    .
\end{equation}

In practice, we performed the reverse transformation and decoupled one of the $o$-solutes from the rest of the system over a series of 21 simulations with equally spaced $\lambda$-values decreasing from 1 to 0. 
We simulated each $\la$-value for 20~ns in the constant NPT ensemble according to the protocols described in Section~\ref{subsubsec-simparam}.
We avoided singularities from simulations with very weakly coupled solutes by employing soft-core LJ potentials with an alpha parameter of 0.5, a sigma parameter of 1 nm, and a soft-core exponent of 1.\cite{BEUTLER1994529}
We then integrated the resulting thermodynamic force with the GROMACS implementation of BAR.\cite{bennett1976}

\section{Results and Discussion}
\label{sec:results_discussion}
Dilute solution theory (DST) provides a simple theory for predicting the influence of dilute solutes upon the interface between coexisting liquid and vapor phases. 
DST treats solute interactions to lowest order and derives the relevant thermodynamic parameters in terms of microscopic partition functions. 
This section numerically investigates DST predictions for dilute solutions with a dominant solvent ($w$) and one or more distinct solute ($o$) species.  
For simplicity and computational efficiency, we model all molecules as spherically symmetric Lennard-Jones (LJ) particles with the same mass and LJ diameter. 
We define the energy scale by setting the depth of the solvent-solvent ($ww$) LJ potential to 1, i.e., $\eps_{ww} =1 $.
Solute species are then distinguished by the LJ well depths describing the solute-solvent ($ow$) and solute-solute ($oo$) interactions.

\begin{table}[h]
\caption{
Microscopic and thermodynamic parameters for binary solutions.
The second and third columns report the well-depth of the solute-solvent ($\eps_{ow}$) and solute-solute ($\eps_{oo}$) potentials for each solute species.
The fourth, fifth, and sixth columns report DST predictions for the corresponding thermodynamic parameters $\ove_{oo} \equiv \bt\ve_{oo}$, $\ko$, and $h_{oo}$, respectively.
\\
}
\label{tab:binary_parameters}
\begin{tabular}{|c|cc|ccc|}        
\hline
Solute  
        &  \hspace{10pt} $\eps_{ow}$ \hspace{5pt}
                & \hspace{5pt} $\eps_{oo}$ \hspace{10pt}
                    & $\ove_{oo}$ 
                        & $k_{o}$ 
                            & $h_{oo}$   
                            \\
   \hline
SA1 
            &  0.90  &  1  &    -5.41    &  1.02   &    6.89  \\
SA2
            &  0.95  &  1  &    -3.07    &  0.41   &    2.27  \\
SA3 
            &  1.05  &  1  &     0.82    & -0.29   &   -0.74  \\
SA4
            &  1.15  &  1  &     3.79    & -0.66   &   0.18  \\
\hline
SN1 
            & 1    &  0.10 &      9.04   & 0       &   -3.60  \\
SN2 
            & 1    &  0.50 &      6.15   & 0       &   -2.56  \\
SN3
            & 1    &  0.80 &      2.41   & 0       &   -1.23  \\
SN4 
            & 1    &  1.20 &     -5.43   & 0       &    1.54  \\
SN5 
            & 1    &  1.50 &    -14.83   & 0       &    4.86  \\
\hline
\end{tabular}
\end{table}

Table~\ref{tab:binary_parameters} presents the LJ parameters for nine different solute species.
We distinguish two classes of $o$-solutes.
In particular, the solute-solvent potentials for surface-neutral (SN) solutes are identical to solvent-solvent potentials, i.e., $\eps_{ow} = \eps_{ww} = 1$. 
Conversely, the solute-solvent potentials for surface-active (SA) solutes differ from solvent-solvent potentials, i.e., $\eps_{ow} \neq \eps_{ww}$. 
Table~\ref{tab:binary_parameters} also reports the DST parameters that were determined for each solute via free energy calculations.
(Note the distinction between the LJ interaction parameter, $\eps_{ij}$, and the thermodynamic parameter, $\ve_{ij}$, describing the influence of interactions upon the free energy.)
Section~\ref{sec:theory} derives these thermodynamic parameters in terms of ratios of partition functions for 1 or 2 solvated solutes, while section~\ref{sec:methods} describes our computational methods for determining these parameters.

\subsection{Excess chemical potentials}
\label{subsec-chempotls}
We first consider the predictions of DST for modeling liquid solutions at the coexistence pressure for pure solvent, $\Pnull$ .
The solute chemical potential, $\muo$, in a binary solution may be decomposed:
\begin{equation}
\label{eq-thermodef}
\muo = \muoid + \muoxs	.
\end{equation}
Here we define the ideal contribution, $\muoid \equiv \muost + \kT \ln \mo$, 
where $\mo = \no/\nw$ is the solute molality.
The excess contribution is then $\muoxs \equiv \muo - \muoid = \veoo \mo + \mO(\mo^2)$, where $\veoo$ is a thermodynamic parameter describing solute-solute interactions at infinite dilution.
DST treats these solute-solute interactions at lowest order, $\muoxs \approx \veoo \mo$, where $\veoo$
is determined by Eq.~\eqref{def-oveloop} according to Hill's constant pressure solution theory.\cite{hillTheorySolutionsI11957}

\begin{figure}[h]
    \centering
    \includegraphics[scale=1.0]{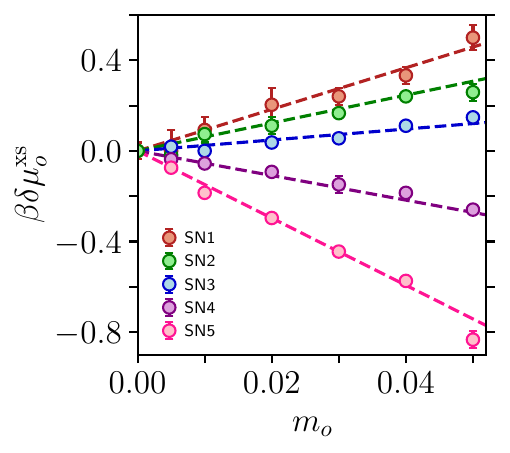}
    \caption{Excess chemical potentials for SN solutes. The symbols report the results of numerical calculations via thermodynamic integration.
    The dashed lines report the DST predictions, $\muoxs \approx \veoo \mo$. 
    }
    \label{fig:excess_mu}
\end{figure}

Figure~\ref{fig:excess_mu} compares the DST approximation for $\muoxs \approx \veoo\mo$ (dashed lines) with numerically exact calculations via thermodynamic integration (filled symbols).
Specifically, Fig.~\ref{fig:excess_mu} considers five surface-neutral solutes with $\eps_{ow} = \eps_{ww}$ that differ in the strength of the solute-solute  potential.

The solute-solute attractions for SN1, SN2, and SN3 solutes are relatively weak with $\eps_{oo}/\eps_{ww}$ = 0.10, 0.50, and 0.80, respectively.
DST predicts the corresponding effective interaction parameters parameters to be $\ove_{oo} \equiv \bt\ve_{oo} = $ 9.04, 6.15, and 2.41, respectively.
These effectively repulsive interactions cause the solute chemical potential to increase with increasing solute concentration.

Conversely, the solute-solute attractions for SN4 and SN5 solutes are relatively strong with $\eps_{oo}/\eps_{ww}$ = 1.20 and 1.50, respectively.
DST predicts the corresponding effective interaction parameters to be $\ove_{oo} =  -5.43$ and $-14.83$, respectively. 
These attractive interactions cause the solute chemical potential to decrease with increasing solute concentration.

DST predicts the solute chemical potentials with nearly quantitative accuracy.
Since DST only treats solute-solute interactions to lowest order, one expects DST will fail for higher concentrations when attractive solutes begin to cluster, as well as for solutes with more attractive interactions.
Nevertheless, DST accurately models these five solutions up to a surprisingly high concentration, $m \le 0.05$.
If the solvent is similar to water with a molarity of approximately 55M, then $m = 0.02$ corresponds to approximately molar solute concentration.

\subsection{Liquid-vapor coexistence}
\label{subsec-coex}
We next investigate DST predictions for liquid-vapor coexistence of dilute binary solutions.
While we have treated solute-solute interactions to lowest order in the liquid phase, we have neglected these interactions in the vapor phase.
Consequently, DST predicts that the solute density in the vapor phase, $\rhovo$, should be related to the molality, $\mo$, of the coexisting solution by
\begin{equation}
\label{eq-coex-results}
\rhovo 
    =   
        \rhostvwz \rKo \mo \left( 1 + \ove_{oo} \mo\right)	
\end{equation}
where $\rhostvwz$ is the density of pure solvent vapor at coexistence and $\rKo$ is a Henry's law constant that is microscopically defined by Eq.~\eqref{def-rKo-maintext}.

\begin{figure}[h]
    \centering
    \includegraphics[scale=1.0]{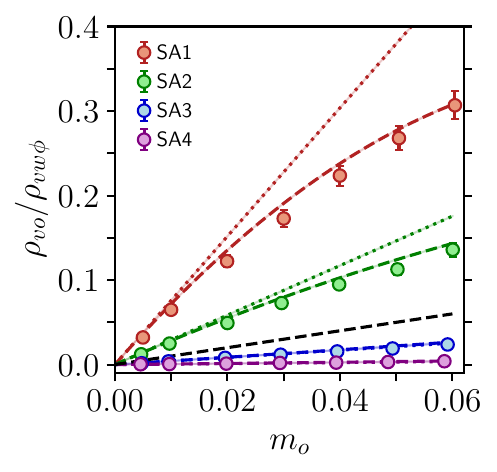}
    \caption{
    Characterization of liquid-vapor coexistence. 
    The dashed curves present DST predictions for the composition, $\rhovo / \rhostvwz$, of the vapor phase as a function of the solution molality, $\mo$, according to Eq.~\eqref{eq-coex-results}.
    The dotted lines present the contribution, $\rKo\mo$, from Henry's law.
    The filled symbols present the results of MD simulations, while the corresponding error bars indicate the standard error.  
    The dashed black line indicates the null case, $\rhovo/\rhostvwz = \mo$.
    }
    \label{fig:henry_law}
\end{figure}

Figure~\ref{fig:henry_law} assesses DST predictions (dashed lines) for the composition, $\rhovo / \rhostvwz$, of the coexisting vapor phase.
Specifically, Fig.~\ref{fig:henry_law} considers four surface-active (SA) solutes for which solute-solute and solvent-solvent interactions are of equal strength, $\eps_{oo} = \eps_{ww}$. 
Instead, SA solutes are distinguished by the strength of their solvent-solute interactions, $\eps_{ow} \neq \eps_{ww}$.
The dashed black line in Fig.~\ref{fig:henry_law} presents the null case, $\rhovo / \rhostvwz = \mo$, for which the coexisting phases have the same composition.

SA1 and SA2 solutes interact with solvent via relatively repulsive LJ potentials with $\eps_{ow}/\eps_{ww}$ = 0.90 and 0.95, respectively.
Because they have relatively weak attraction to solvent, these solutes  preferentially partition into the vapor phase. 
Table~\ref{tab:Ko_vals} reports DST predictions of the Henry's law constant $\rKo = $ 7.59 and 2.93 
for SA1 and SA2 solutes, respectively.
Henry's law (dotted lines) accurately describes this liquid-vapor coexistence up to modest concentrations, $\mo \le 0.01$.
At higher concentrations, though, attractive solute-solute interactions ($ \ove_{oo} < 0$)  in the liquid solution reduce the preference of these solutes for the vapor phase. 
DST very accurately captures this effect up to rather high concentrations, $\mo \le 0.06$.

\begin{table}
\caption{
DST predictions of the Henry's law constant, $\rKo$, for SA solutes.
}
\label{tab:Ko_vals}
\begin{tabular}{|cc|}
\hline
Solute  & $\rKo$ \\
   \hline
SA1 
            & 7.59     \\
SA2 
            & 2.93     \\
SA3 
            & 0.42     \\
SA4 
            & 0.06     \\
\hline
\end{tabular}
\end{table}

Conversely, SA3 and SA4 solutes interact with solvent via relatively attractive LJ potentials with $\eps_{ow}/\eps_{ww}$ = 1.05 and 1.15, respectively.
Consequently, these solutes are significantly depleted from the vapor phase.
Table~\ref{tab:Ko_vals} reports DST predictions of the Henry's law constant $\rKo = $ 0.42 and 0.06 for SA3 and SA4, respectively.
Henry's law accurately describes liquid-vapor coexistence for these solutes at all concentrations, $\mo \le 0.06$.
In this case, the effective repulsion between solvated solutes has little effect upon  their partitioning between the liquid and vapor phases. 
Thus, DST very accurately describes liquid-vapor coexistence in all cases that we considered.

\subsection{Interfacial properties of dilute binary solutions}
\begin{figure}[h]
    \centering
    \includegraphics[scale=1.0]{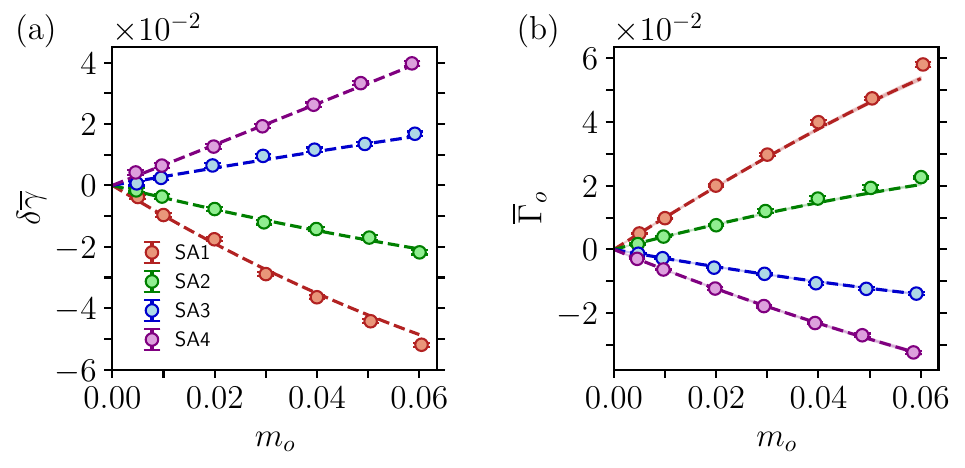}
    \caption{
    Characterization of the liquid-vapor interface for binary solutions with surface active solutes.
    Panels (a) and (b) report the surface tension and surface excess, respectively. 
    Dashed curves report DST predictions, while symbols report the results of detailed MD simulations.
    Error bars represent one standard error in the simulated results.
    }
    \label{fig:k_neq_zero}
\end{figure}

\begin{figure}[h]
    \centering
     \includegraphics[scale=1.0]{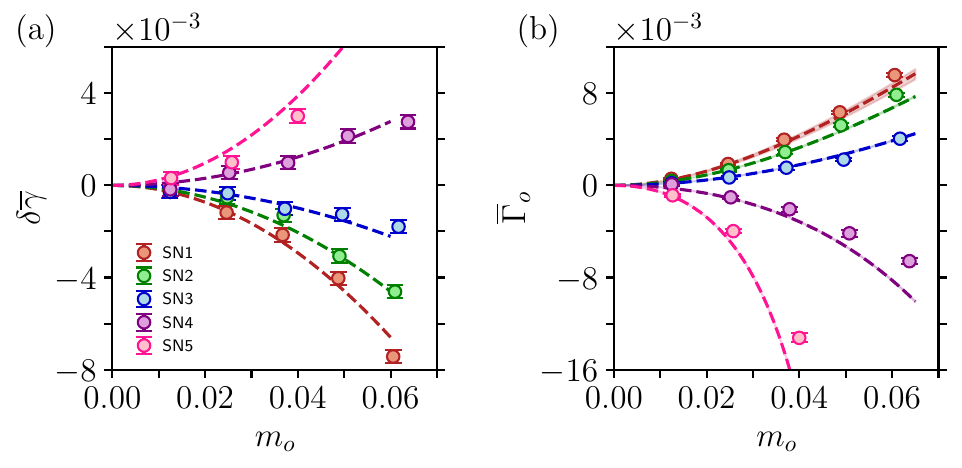}
    \caption{
        Characterization of the liquid-vapor interface for binary solutions with surface neutral solutes. 
    Panels (a) and (b) report the surface tension and surface excess, respectively. 
    Dashed curves report DST predictions, while symbols report the results of detailed MD simulations.
    Error bars on the points represent one standard error in the simulated results. 
    }
    \label{fig:k_zero}
\end{figure}

We now consider DST predictions for the interfacial properties of dilute binary solutions.
DST predicts that the surface tension and surface excess should vary with solute concentration according to 
\begin{eqnarray*}
\dt\bga 
		& \approx &  
	 		- 
				\ko \mo 
			+
				\half h_{oo} \mo^2 
								\\
\bGao 
		& \approx & 
		\mo
		\left( 
			\frac{\ko- h_{oo} \mo}{1+\ove_{oo} \mo}
		\right)
		,
\end{eqnarray*}
where $\ko$ is the intrinsic interfacial preference and $h_{oo}$ describes solute-solute interactions at the interface.
DST predicts $\ko$ and $h_{oo}$ in terms of microscopic partition functions according to Eqs.~\eqref{def-ko} and \eqref{def-hoop}, respectively.
Figures~\ref{fig:k_neq_zero} and \ref{fig:k_zero} compare DST predictions (dashed curves) with the results (filled symbols) of MD simulations.

Figure~\ref{fig:k_neq_zero} considers the surface-active solutes that were introduced in Section~\ref{subsec-coex}.
Because they have relatively little attraction for solvent molecules, SA1 and SA2 solutes behave as weak surfactants with positive intrinsic interfacial preferences, $\ko = 1.02$ and 0.41, respectively. 
However, attractive effective interactions with solutes in the bulk liquid phase ($\ove_{oo} < 0$) slightly reduce this interfacial preference at higher concentrations.  
Conversely, because of their relatively strong attraction to solvent, SA3 and SA4 solutes behave as weak depletants with negative intrinsic interfacial preference, $\ko = -0.29$ and $-0.66$, respectively. 
In this case, repulsive effective interactions with solutes in the bulk ($\ove_{oo} > 0$) slightly reduce their preference for the bulk liquid at higher solute concentrations. 
DST predicts the interfacial properties for these surface-active solutes with nearly quantitative accuracy over the entire concentration range.

Figure~\ref{fig:k_zero} considers the surface-neutral solutes that were introduced in Section~\ref{subsec-chempotls}.
Because their interactions with solvent are equivalent to solvent-solvent interactions, surface-neutral solutes do not demonstrate an intrinsic interfacial preference, $\ko = 0$.
Consequently, in comparison to the surface-active solutes in Fig.~\ref{fig:k_neq_zero}, surface neutral solutes exert a significantly smaller and more subtle influence upon interfacial properties that is primarily governed by bulk solute-solute interactions.
SN1, SN2, and SN3 solutes act as weak effective surfactants with a slight  preference for the interface in order to minimize the effective repulsion ($\ove_{oo} > 0$) with other solutes in the liquid solution.
Conversely, SN4 and SN5 solutes act as weak effective depletants with a slight preference for the bulk in order to maximize the effective attraction ($\ove_{oo} < 0$) with other solutes in the liquid solution.
Notwithstanding the discrepancies for SN5 solutes at relatively high concentrations, DST predicts the surface tension and surface excess for these binary solutions quite accurately.

\subsection{Surfactant to depletant transition in ternary solutions}
Finally, we consider a ternary solution with two dilute cosolutes, $o = $ 1 and 2.
DST predicts that the presence of cosolute 2 can significantly alter the interfacial preferences of cosolute 1.
For instance, attractive cross-interactions can convert weak intrinsic surfactants into weak effective depletants.
Our previous study illustrated this transition for a simple lattice model of solutions.\cite{mandalaparthy2022a}
We now examine this transition in a more realistic off-lattice model that also explicitly accounts for liquid-vapor coexistence.

\newcommand{\veps}{\varepsilon}
\begin{table}[h]
\caption{Microscopic and thermodynamic parameters for symmetric ternary solution. 
}
\label{tab:ternary_parameters}
\begin{tabular}{|cc|}
\hline
Parameter  & Value  \\
   \hline
$\eps_{ww}$ & 1.0 \\
$\eps_{1w} = \eps_{2w}$ & 0.99 \\
$\eps_{11} = \eps_{22}$ & 1.0 \\
$\eps_{12}$ & 1.5 \\
\hline
$\ove_{11} 
    =  \ove_{22}$ 
        & -1.39\\
$k_{1}  
    = k_{2}$  
        & 0.067\\
$h_{11}
    = h_{22}$   
        & 0.28 \\
$\ove_{12}$   
        & -15.56\\
$h_{12}$    
        & 5.91 \\
\hline
\end{tabular}
\end{table}
Specifically, we consider the symmetric ternary solution that is described by Table~\ref{tab:ternary_parameters}.
The attraction of each solute for the solvent is slightly weaker than the solvent-solvent attraction, $\eps_{1w} = \eps_{2w} = 0.99 \eps_{ww}$.
Consequently, DST predicts that each solute should behave as a weak intrinsic surfactant with a very slight intrinsic interfacial preference, $k_1 = k_2 = 0.067.$ 
The (self) solute-solute interaction is equivalent to the solvent-solvent interaction for both cosolutes, i.e., $\eps_{11} = \eps_{22} = \eps_{ww}$, such that $\veps_{11} = \veps_{22} = -1.39 \kT$.
Conversely, the (cross) attraction between different cosolute species is 50\% stronger than the solvent-solvent attraction, i.e., $\eps_{12} = \eps_{21} = 1.50 \eps_{ww}$, such that $\veps_{12} = \veps_{21} = -15.56\kT$.

\begin{figure}[h]
    \centering
    \includegraphics[scale=1.0]{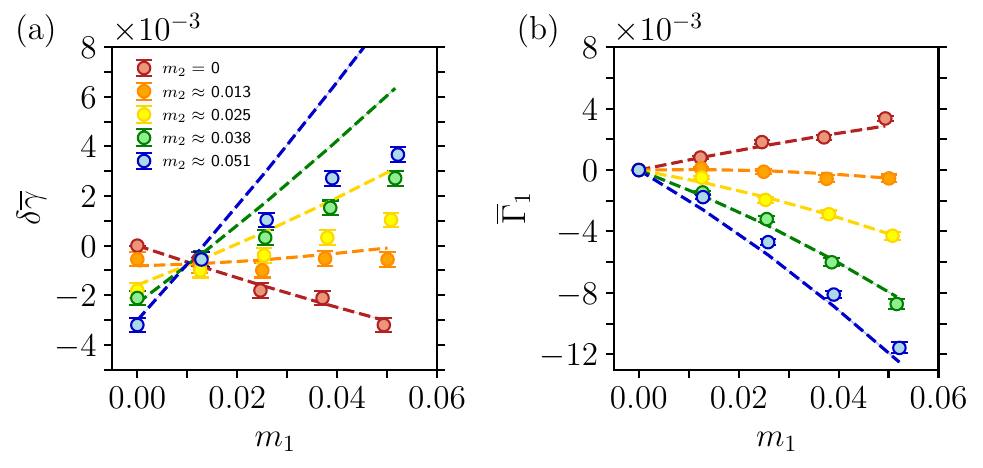}
    \caption{
    Characterization of the liquid-vapor interface for ternary solutions.
    Panels (a) and (b) report the surface tension and surface excess, respectively, as a function of the solute 1 molality, $m_1$.
    The colors indicate the concentration of solute 2 in each solution.
    Dashed curves report DST predictions, while symbols report the results of detailed MD simulations.
    Error bars report one standard error in the simulated results.
    }
    \label{fig:ternary}
\end{figure}

Figure~\ref{fig:ternary} compares DST predictions (dashed curves) with the results of numerically exact MD simulations (filled symbols).
Each color corresponds to solutions with (approximately) constant concentration of cosolute 2.
In the absence of cosolute 2 (red, $m_2 = 0$), cosolute 1 does indeed behave as a weak surfactant, as it slightly reduces the surface tension and slightly accumulates at the interface. 
Since the two cosolutes are equivalent, the solutions corresponding to $m_1 = 0$ in Fig.~\ref{fig:ternary}a demonstrate that cosolute 2 has the same influence on the interface in the absence of cosolute 1.
As cosolute 2 is added, the attractive cross-interactions significantly reduce the interfacial preference of cosolute 1. 
When $m_2 \approx 0.013$, cosolute 1 demonstrates essentially no preference for the interface and exerts minimal influence upon the surface tension.
At higher concentrations of cosolute 2, cosolute 1 behaves as an effective depletant and increases the surface tension. 
Thus, the addition of cosolute 2 converts cosolute 1 from a weak intrinsic surfactant into a weak effective depletant. 
Conversely, Fig.~\ref{fig:ternary}a demonstrates that the addition of cosolute 2 reduces the surface tension when $m_1 < 0.01$, but increases the surface tension when $m_1 > 0.01$.
There is a critical cross-over point near $m_1 \approx 0.01$ where the addition of cosolute 2 has no influence upon the surface tension.

DST predicts the simulated interfacial properties with reasonable accuracy. 
In particular, DST predicts the surface tension and surface excess with nearly quantitative accuracy for $m_1 + m_2 \le 0.04$. 
Moreover, DST very accurately predicts the surface excess at all simulated concentrations. 
At higher concentrations, DST slightly overestimates the surface tension when $m_2 \ge 0.025$ and $m_1 + m_2 \ge 0.05$.
We suspect that this discrepancy arises from cooperative many-body interactions in the liquid phase that occur as the cosolutes begin to  aggregate. 
Notwithstanding these discrepancies at  relatively high concentrations, DST predicts the interfacial properties with rather satisfactory accuracy.
Most importantly, DST correctly predicts the qualitative transition of the cosolutes from weak intrinsic surfactants into weak effective depletants.

\section{Conclusions}
\label{sec:conclusions}
Dilute solution theory (DST) provides a simple framework for understanding the influence of dilute solutes upon liquid interfaces.\cite{mandalaparthy2022,mandalaparthy2022a} Importantly, it derives the relevant thermodynamic parameters in terms of microscopic partition functions. This derivation not only connects these parameters to molecular interactions, but also provides a computational framework for predicting these parameters. Because it treats solute-solute interactions at lowest order, in some sense, DST provides the simplest rigorous theory for understanding how these interactions can modulate interfacial properties. In particular, DST highlights the distinction between intrinsic and effective interfacial preferences. While the intrinsic preference reflects the interactions of infinitely dilute solutes with solvent, the effective preference reflects solute-solute interactions in both bulk and interfacial environments. For instance, attractive bulk interactions can convert weak intrinsic surfactants into weak effective depletants. 

The present work makes several important advances to this DST. While our previous work relied upon lattice models, the present work treats off-lattice models that interact via much more realistic potentials. More importantly, the present work explicitly accounts for coexistence between fluid phases. Intriguingly, the Gibbs ensemble\cite{Panagiotopoulos01071987,panagiotopoulosPhaseEquilibriaSimulation1988} provides an ideal reference state for isolating interfacial properties. Additionally, we have developed approximations for determining the relevant bulk contributions based upon simpler simulations of bulk phases. While we focused on liquid-vapor interfaces in this work, we anticipate that the present derivation should readily generalize to more complex fluid interfaces and, in particular, coexisting liquid phases. Consequently, we anticipate that the present framework may prove fruitful for understanding the influence of cosolutes upon the interface between coexisting liquid domains in biological systems.\cite{berryPhysicalPrinciplesIntracellular2018}  

The present work also indicates several directions for further developing DST. In particular, it may be beneficial to more clearly link macroscopic interfacial properties and specific molecular interactions. As already noted above, DST directly relates the relevant thermodynamic parameters to ratios of microscopic partition functions. While rigorous, this approach is not microscopically transparent because these partition functions convolute many diverse molecular interactions. It may be possible to re-express these partition function ratios as free energy calculations along appropriate reaction coordinates. The resulting free energy profiles may facilitate both numerical calculations and physical insight. Alternatively, it may be possible to develop simplified, approximate treatments that directly relate these partition function ratios to molecular interactions. For instance, the intrinsic interfacial preference, $k_o$, is the ratio of effective partition functions for a single solute at the interface and in the bulk liquid. It may be useful to develop phenomenological or mean field expressions for these effective partition functions that, e.g., consider the influence of the interface upon the solute solvation shell. Similarly, further analysis of DST for lattice models may clarify the relationship between thermodynamic and molecular parameters. 

Future studies should certainly investigate DST for more realistic models of, e.g., aqueous osmolyte solutions. Similarly, future work should extend DST to treat salt solutions and other systems with charged solutes. Moreover, in its present form, DST predictions are clearly limited by the accuracy of the underlying microscopic model, which may be particularly problematic for interfacial phenomena. Consequently, it may be beneficial to re-express DST parameters in terms of experimental observables.\cite{ben-naimInversionKirkwoodBuff1977} For instance, our previous work suggested that the surface tension and surface excess for molar osmolyte solutions may be predicted from vapor pressure osmometry measurements and surface tension measurements at infinite dilution.\cite{mandalaparthy2022a,autonMetricsThatDifferentiate2006,courtenayVaporPressureOsmometry2000}

More fundamentally, DST employs a perturbation theory in solute concentration about a reference state that, in this case, corresponds to pure solvent. It should be possible to extend this perturbation theory to more complex reference states that include, e.g., a fixed concentration of proteins. Similarly, it may be fruitful to generalize DST in analogy to Kirkwood-Buff theory,\cite{kirkwoodStatisticalMechanicalTheory1951,BenNaim:2006,ploetzLocalFluctuationsSolution2013} although one expects that such approaches may fail when solutes begin to aggregate. Nevertheless, we hope that the present work may prove useful for modeling and interpreting the influence of dilute solutes upon liquid interfaces.

\section*{Supplementary Material}
See the supplementary material for a more detailed derivation of the off-lattice theory and a more detailed discussion of our computational methods.

\begin{acknowledgments}
The authors gratefully acknowledge helpful comments by Athanassios  Panagiotopoulos, Nico van der Vegt, and Adam Willard. 
Portions of this research were conducted on resources provided by Prof. Nico van der Vegt (TU Darmstadt).
Portions of this research were conducted with Advanced CyberInfrastructure computational resources provided by The Institute for Computational and Data Sciences at The Pennsylvania State University (http://icds.psu.edu).
Additionally, parts of this research used the Expanse resource at the San Diego Supercomputer Center through allocation TG-CHE170062 from the Extreme Science and Engineering Discovery Environment (XSEDE)\cite{towns2014}, which was supported by National Science Foundation grant number TG-CHE170062.
This work also used allocation CHE170062 from the Advanced Cyberinfrastructure Coordination Ecosystem: Services \& Support (ACCESS) program, which is supported by National Science Foundation grants \#2138259, \#2138286, \#2138307, \#2137603, and \#2138296.
Portions of this work were carried out at the Advanced Research Computing at Hopkins (ARCH) core facility  (rockfish.jhu.edu), which is supported by the National Science Foundation (NSF) grant number OAC1920103.
VM thanks Taylor Zaniewski for her assistance in composing Fig 1.
We employed Matplotlib for all figures.\cite{hunter2007}
\end{acknowledgments}

\section*{Author Declarations}
\subsection*{Conflict of interest}
The authors have no conflicts to disclose.

\section*{Data Availability}
The data that support the findings of this study are available from the corresponding author upon reasonable request.

\bibliography{Off_Lattice_DST}

\end{document}


\title{Supplementary Material for: Relating solute interactions to interfacial properties}

\author{Varun Mandalaparthy}
\author{Benjamin M.~Curlee}
\affiliation{Department of Chemistry, The Pennsylvania State University,  University Park, Pennsylvania 16802, USA}
\author{W.~G.~Noid}
\email{wnoid@chem.psu.edu}
\affiliation{Department of Chemistry, The Pennsylvania State University,  University Park, Pennsylvania 16802, USA}
\date{\today}

\maketitle

\section{Detailed Theory}
\subsection{Coexisting phases}
\label{Sec-Theory-Coexist}
\subsubsection{Thermodynamics of interfaces}
\label{SubSec-Theory-CoexThermo}
We consider an interfacial system of $\Nw$ solvent ($w$) molecules and $M$ different types of dilute $o$-solutes in a fixed volume $V$ at a temperature, $T$, as illustrated in Fig.~1 of the main text. 
We assume that the interfacial system lies in the 2-phase liquid-vapor region of the phase diagram and is far from the critical point.
The system phase separates into homogeneous liquid and vapor regions that are separated by an interface with an area, $\si$.
Coexistence requires that the liquid and vapor regions must have the same pressure and the same chemical potentials:
\begin{equation}
\label{eq-coexist}
\begin{array}{lcl}
P_l(\rholw,\setrholo) 		& = & P_v(\rhovw,\setrhovo)		= P 		\\
\mulw(\rholw,\setrholo) 	& = & \muvw(\rhovw,\setrhovo)		= \muw	\\
\mulo(\rholw,\setrholo) 	& = & \muvo(\rhovw,\setrhovo)		= \muo  \hspace{0.5in}
											\text{for $o = 1,\ldots, M$},
\end{array}
\end{equation}
where, e.g., $\rholw$ and $\rholo$ indicate the density of solvent and $o$-solutes, respectively in the liquid region. 
Throughout this work, we assume $T$ is a constant and treat it implicitly.

We select a Gibbs dividing surface that is equimolar with respect to the solvent.
Specifically, we define the volume of the liquid, $V_l$, and vapor, $V_v$, regions such that 
\begin{equation}
\label{def-GibbsSurface}
\begin{array}{lcl}
V 		& = & V_l + V_v	\\
N_w 	& = & V_l \rholw + V_v \rhovw 	 
\end{array}		.
\end{equation}
In contrast, the total number of $o$-solutes, $\Nto$, cannot be accounted for by the two bulk phases:
\begin{equation}
\Nto = \Nbo + \Nio	,
\end{equation}
where the bulk number of $o$-solutes is 
\begin{equation} 
\label{def-Nbo}
\Nbo = V_l \rholo + V_v \rhovo	, 
\end{equation}
while $\Nio$ accounts for the net accumulation or depletion of solutes near the interface.
The Helmholtz potential for the total inhomogeneous system, $\At$, may be similarly decomposed
\begin{equation}
\At = \Ab + \Ai
\end{equation}
where the bulk contribution is 
\begin{eqnarray}
\label{def-Ab}
\Ab 	& = & 
			\Ab(\setNbo) 
		= 
			\Al(\Nlw, \setNlo, \Vl) + \Av(\Nvw, \setNvo, \Vv) 	\\
\label{eq-Ab}
	& = & 
		-     P V + \muw \Nw + \sumo \muo \Nbo
\end{eqnarray}
and the interfacial contribution is 
\begin{equation}
\Ai = \left\{ \ga + \sumo \muo \Gao \right\} \si
\end{equation}
where $\ga$ is the surface tension and $\Gao = N_\io / \si$ is the surface excess of $o$-solutes.
The Gibbs adsorption equation relates this surface excess to the variation in the surface tension (at constant $T$):
\begin{equation}
\label{eq-Gibbs-adsorption-eqn}
\dd\ga = - \sumo \Gao \dd\muo	.
\end{equation}

The coexistence criterion in Eq.~\eqref{eq-coexist} give $2+M$ equations for $2 + 2M$ densities. 
Consequently, the liquid phase densities for the $M$ $o$-solute species, $\setrholo$, completely determine the composition of the two coexisting phases. 
Given fixed $\Nw$ and $V$, the densities then determine the volumes of the liquid and vapor phases from Eq.~\eqref{def-GibbsSurface}.
This then determines $\setNbo$ from Eq.~\eqref{def-Nbo}.
Thermodynamic reasoning suggests that this argument should also work in reverse.
Specifically, the $M+2$ `bulk' extensive variables $(\Nw, \setNbo, V)$ and the coexistence criterion of Eq.~\eqref{eq-coexist} should determine the composition and extent of the two coexisting phases.
This follows because, given fixed bulk values $(\Nw, \setNbo, V)$, Eqs.~\eqref{def-GibbsSurface} and \eqref{def-Nbo} allow us to express the $M+2$ coexistence criterion in terms of $M+2$ variables, e.g., $(\rholw, \setrholo, \Vl)$.
We assume this system of equations will have a unique solution in the two phase region for dilute solutions.

\subsubsection{Closed ``bulk'' ensemble}
\label{Subsec-Theory-ClosedBulkEnsemble}
We return to Fig.~1 of the main text and consider two spatially separated subsystems that are far from the liquid-vapor interface.
The ``vapor'' subsystem contains $\Nvw$ solvent molecules and $\setNvo$ solute molecules in a volume, $\Vv$, that is located in the bulk vapor region.
The ``liquid'' subsystem contains $\Nlw$ solvent molecules and $\setNlo$ solute molecules in a volume, $\Vl$, that is located in the bulk liquid region.

We define a closed ``bulk'' ensemble by allowing the volumes and compositions of the two subsystems to fluctuate, but requiring that the total volume and the total number of each species are fixed, i.e., 
\begin{equation}
\label{eq-bulk-constr}
\begin{array}{lcl}
\Vv	 + \Vl 	& = & V 	= \text{const}		\\
\Nvw	 + \Nlw 	& = & \Nw = \text{const}		\\
\Nvo + \Nlo 	& = & \Nbo = \text{const} \hspace{0.5in} \text{for $o = 1, \ldots, M$}.
\end{array}
\end{equation}
The partition function for this closed bulk ensemble is 
\begin{eqnarray}
\nonumber
\Qb(\setNbo) 
	& \equiv &  
		\Qb(\Nw, \setNbo,V)
	\\
\label{def-Qb}
	& \equiv &  
		\sum_{\Vl} \sum_{\Nlw}  \sum_{\setNlo} 
			\Ql(\Nlw,\setNlo,\Vl) 
			\Qv(\oNvw,\{\oNvo\},\oVv)	,
\end{eqnarray}
where $\Ql$ and $\Qv$ are conventional canonical partition functions, while $\oNvw = \Nw - \Nlw$, $\oNvo = \Nbo - \Nlo$, and $\oVv = V - \Vl$ are vapor variables determined by the values of liquid variables and the fixed constraints. 
This closed bulk ensemble corresponds to the Gibbs ensemble for modeling two-phase coexistence without simulating explicit interfaces.
We now show that this Gibbs ensemble corresponds to the bulk regions of the inhomogeneous total system considered in Section~\ref{SubSec-Theory-CoexThermo}.

\newcommand{\mPb}{\mP_b}

The probability, $\mPb$, that the liquid region contains $\Nlw$ solvent molecules and $\setNlo$ solute molecules in a volume $\Vl$ is given by
\begin{eqnarray}
\label{def-Pb}
\mPb(\Nlw,\setNlo,\Vl) 
	& \equiv & 
		\Ql(\Nlw,\setNlo,\Vl) \times \Qv(\oNvw,\{\oNvo\},\oVv)	 / \Qb(\setNbo)		\\
\label{eq-Pb}
	& = & 
		\exp\left[
			-\bt 
			\left( 
				\Al(\Nlw, \setNlo, \Vl) + \Av(\oNvw, \{\oNvo\}, \oVv) 
			\right) 
			\right] / \Qb(\setNbo)	.
\end{eqnarray}
For large systems, we expect that $\mPb$ will be sharply peaked about the most probable partitioning, $(\Nlw',\{\Nlo'\},\Vl')$.
By differentiating Eq.~\eqref{eq-Pb}, while accounting for the fact that intensive quantities depend only upon intensive properties, we find that the most probable partitioning corresponds to the following system of equations:
\begin{equation}
\label{eq-max-cond}
\begin{array}{lcl}
P_l(\rholw',\{\rholo'\}) 	& = & 	P_v(\rhovw',\{\rhovo'\})	= P'		\\
\mulw(\rholw',\{\rholo'\}) 	& = & \muvw(\rhovw',\{\rhovo'\})		= \muw'	\\
\mulo(\rholw',\{\rholo'\}) 	& = & \muvo(\rhovw',\{\rhovo'\})		= \muo'  \hspace{0.5in}
											\text{for $o = 1,\ldots, M$}.
\end{array}
\end{equation}
Thus, as is well known, the two regions of the Gibbs ensemble fluctuate about coexisting liquid and vapor phases.

According to the argument at the end of Subsection~\ref{SubSec-Theory-CoexThermo}, the variables defining the closed bulk ensemble, $(\Nw,\setNbo,V)$, along with the coexistence conditions of Eq.~\eqref{eq-max-cond} correspond to a unique set of solute chemical potentials.
Thus, if we define the closed bulk ensemble variables
to match the interfacial system, then the liquid and vapor regions sampled by the closed bulk ensemble 
correspond to the liquid and vapor regions of the interfacial system that was considered in Sec.~\ref{SubSec-Theory-CoexThermo}.
Moreover, in the limit that the closed bulk ensemble is sufficiently large, we expect that 
\begin{equation}
\Qb(\setNbo) \approx \Ql(\Nlw',\{\Nlo'\},\Vl') \times \Qv(\Nvw',\{\Nvo'\},\Vv')	,
\end{equation}
such that the Helmholtz potential for the closed bulk ensemble equals the bulk contribution to the Helmholtz potential for the interfacial system, i.e., $\Ab$ in Eq.~\eqref{def-Ab}.
Consequently, the closed bulk ensemble can be used to subtract off the bulk contributions to the Helmholtz potential for the interfacial system.

\subsection{Statistical mechanics of interfacial properties}
\label{Sec-StatmechIntfaclProp}
\subsubsection{Semi-grand ensemble for inhomogeneous system}
\label{SubSec-Theory-SemiGrandTotal}
We now consider the statistical mechanics of the inhomogeneous ``total'' system introduced in Sec.~\ref{SubSec-Theory-CoexThermo}.
The total system has a fixed number of solvent molecules, $\Nw$, as well as a fixed volume, $V$, temperature, $T$, and interfacial area, $\si$.
In the absence of solutes, the Helmholtz free energy, $\Atnull$, and canonical partition function, $\Qtnull$, for the solvent are 
\begin{equation}
\Atnull = -\kT \ln \Qtnull = -\Pnull V + \muwnull \Nw + \ganull \si 		,
\end{equation}
where $\Pnull$, $\muwnull$, and $\ganull$ indicate the pressure, solvent chemical potential, and surface tension of pure solvent in the absence of $o$-solutes.
In the presence of $\setNto$ $o$-solutes, this becomes 
\begin{equation}
\At(\setNto) = - \kT \ln \Qt(\setNto) = - P V + \muw \Nw + \sumo \muo \Nto + \ga \si.
\end{equation}
We denote $\Qto$ as the canonical partition function when the inhomogeneous system contains a single $o$-solute.
Similarly, we denote $\Qtoop$ as the canonical partition function when the inhomogeneous system contains exactly two solutes that are of type $o$ and $o'$.

We now suppose that the inhomogeneous total system is open to $o$-solutes at constant chemical potentials, $\setmuo$, while $\Nw$, $V$, $T$ and $\si$ all remain constant.
The relevant free energy for this semi-grand ensemble is 
\begin{equation}
\tAt(\setmuo) 
	= \At(\setNto) - \sumo\muo\Nto 
	= -    P V + \muw    \Nw + \ga \si		.
\end{equation}
For variations at constant $\Nw$, $V$, and $T$, 
\begin{equation}
\dd\tAt 
	= 
		- \sumo \Nto    \: \dd\muo + \ga \:  \dd\si		.
\end{equation}
The relevant partition function is 
\begin{eqnarray}
\tQt(\setmuo) 
	& \equiv &  
		e^{-\bt \tAt}
	\equiv 
		\sum_{\setNtoZero}	\Qt(\setNto) \prod_o \lao^\Nto			\\
 	 & = & 
		\Qtnull + \sumo \Qto \lao + \half \sumoop \lao \Coop \Qtoop \la_\op + \mO(\lao^3),
\end{eqnarray}
where $\lao = e^{\bt \muo}$ is the absolute activity of $o$-solutes and the factor $\Coop = 1 + \dt_{\oop}$ accounts for indistinguishable solutes.

We now define one- and two-solute scaled partition functions:
\begin{eqnarray}
\qto 		& \equiv & 	\Nw\inv \Qto / \Qtnull		\\
Z_{t\oop}	& \equiv & 	\Coop \qto\inv \qtop\inv Q_{t\oop} / \Qtnull		,
\end{eqnarray}
as well as an activity for the inhomogeneous system 
\begin{equation}
\ato \equiv \qto \lao		.
\end{equation}
In order to isolate the contributions from $o$-solutes, we define
\begin{eqnarray}
\label{def-Psit}
\Psi_t(\setmuo) 
	& \equiv & 
			\Qtnull\inv \tQt(\setmuo) = e^{-\bt \dt \tAt}	\\
	& = & 	
			1 
			+ \Nw \sumo \ato 
			+ \half \sumoop \ato Z_{t \oop} \atop + \mO(\ato^3) 	.
\end{eqnarray}
Here, we have defined 
\begin{equation}
\dt\tAt 
	\equiv 
		\tAt(\setmuo) - \Atnull 
	= 
		- V \dt  P    + \Nw\dt\muw    + \si \dt\ga	,
\end{equation}
which reflects the influence of the $o$-solutes upon the coexistence pressure, $\dt P = P - \Pnull$, the solvent chemical potential, $\dt \muw = \muw - \muwnull$, and the surface tension, $\dt \ga = \ga - \ganull$.
From Eq.~\eqref{def-Psit} we obtain an expansion for the free energy in terms of the solute activity:
\begin{equation}
\label{eq-lnPsit}
-\Nw\inv \bt\dt\tAt 
	= \Nw\inv \ln \Psi_t 
	= \sumo \ato - \half \sumoop \ato \ovetoop \atop + \mO(\ato^3) 
\end{equation}
where 
\begin{equation}
\label{def-ovetoop}
\ovetoop \equiv - \Nw\inv \left( Z_{t \oop} - \Nw^2 \right) 	
\end{equation}
is the thermodynamic interaction parameter describing the total system and is analogous to the second virial coefficient for dilute gases or dilute solutions.
If we continued the standard perturbation treatment, we would relate $\ato$, to the ``total molality'' $\mto = \Nto / \Nw$ of the inhomogeneous system and derive an expression for $\dt\tAt$ in terms of $\mto$. 
However, our goal is to determine an expression for surface properties in terms of the molality $\mlo$ of the liquid phase.
Accordingly, we next eliminate the bulk contributions from Eq.~\eqref{eq-lnPsit}.

\newcommand{\Abnull}{A_{b\phi}}

\subsubsection{Semi-grand ensemble for bulk system}
We now perform the same steps for the corresponding bulk system with fixed $(\Nw,V,T)$ that was introduced in Sec.~\ref{Subsec-Theory-ClosedBulkEnsemble}. 
In the absence of solutes, the Helmholtz free energy, $\Abnull$, and canonical partition function, $\Qbnull$ of the bulk system are 
\begin{equation}
\Abnull = -\kT \ln \Qbnull = -\Pnull V + \muwnull \Nw  		.
\end{equation}
In the presence of $\setNbo$ $o$-solutes, this becomes 
\begin{equation}
\Ab(\setNbo) = - \kT \ln \Qb(\setNbo) = - P V + \muw \Nw + \sumo \muo \Nbo .
\end{equation}
Thus, the free energy of the bulk system with $\setNbo$ $o$-solutes equals the bulk contribution to the free energy of the corresponding total interfacial system.

We now consider a semi-grand bulk system that is open to solutes with fixed $\setmuo$.
The relevant free energy is 
\begin{equation}
\tAb(\setmuo) 
	= \Ab(\setNbo)    - \sumo \muo \Nbo 
	= - P    V + \muw    \Nw	.
\end{equation}
For variations at constant $\Nw$, $V$, and $T$, 
\begin{equation}
\dd\tAb 
	= 
		- \sumo \Nbo    \: \dd\muo		.
\end{equation}
The relevant partition function is 
\begin{eqnarray}
\tQb(\setmuo) 
	& \equiv &  
		e^{-\bt \tAb}
	\equiv 
		\sum_{\setNboZero}			\Qb(\setNbo) \prod_o \lao^\Nbo			\\
 	 & = & 
		\Qbnull + \sumo \Qbo \lao + \half \sumoop \lao \Coop \Qboop \la_\op + \mO(\lao^3),
\end{eqnarray}
where $\Qbnull$, $\Qbo$, and $\Qboop$ are canonical partition functions for the bulk system with 0, 1, and 2 solutes, respectively.

We now define one- and two-solute scaled partition functions for the bulk system:
\begin{eqnarray}
\label{def-qbo}
\qbo 		& \equiv & 	\Nw\inv \Qbo / \Qbnull		\\
\label{def-Zboop}
Z_{b\oop}	& \equiv & 	\Coop \qbo\inv \qbop\inv Q_{b\oop} / \Qbnull		,
\end{eqnarray}
as well as a corresponding activity: 
\begin{equation}
\abo \equiv \qbo \lao		.
\end{equation}
We now isolate the solute contributions to the bulk system by defining
\begin{eqnarray}
\label{def-Psib}
\Psi_b(\setmuo) 
	& \equiv & 
			\Qbnull\inv \tQb(\setmuo) = e^{-\bt \dt \tAb}	\\
	& = & 	
			1 
			+ \Nw \sumo \abo 
			+ \half \sumoop \abo Z_{b \oop} \abop + \mO(\abo^3) 	,
\end{eqnarray}
where
\begin{equation}
\dt\tAb 
	\equiv 
		\tAb(\setmuo) - \Abnull 
	= 
		- V \dt   P   + \Nw\dt\muw .
\end{equation}
From Eq.~\eqref{def-Psib} we obtain an expansion for the bulk free energy in terms of the bulk solute activity:
\begin{equation}
\label{eq-lnPsib}
- \Nw\inv \bt\dt\tAb 
	= \Nw\inv \ln \Psi_b 
	= \sumo \abo - \half \sumoop \abo \oveboop \abop + \mO(\abo^3) 
\end{equation}
where 
\begin{equation}
\label{def-oveboop}
\oveboop \equiv - \Nw\inv \left( Z_{b \oop} - \Nw^2 \right) 	.
\end{equation}

\newcommand{\rto}{r_{to}}
\newcommand{\rtop}{r_{t\op}}

\subsubsection{Interfacial properties}
We now eliminate the bulk properties to determine the influence of solutes upon the interfacial free energy
\begin{equation}
\dt\tAi 
	\equiv \dt\tAt(\setmuo) - \dt\tAb(\setmuo) 
	= \si \dt\ga 
	= -\kT \ln \left[\Psi_t/\Psi_b\right]		.
\end{equation}
We use the expansions of Eqs.~\eqref{eq-lnPsit} and \eqref{eq-lnPsib} to obtain 
\begin{equation}
-\Nw\inv \bt \dt\tAi 
	= 
		\sumo (\rto - 1) \abo 
	- 
		\half \sumoop \abo \left[ \rto \ovetoop \rtop - \oveboop \right] \abop 	
	+ 
		\mO( \abo^3 ) ,
\end{equation}
where we have defined
\begin{equation}
\label{def-rto}
\rto = \qto / \qbo 
\end{equation}
and used $\ato = \qto \lao = \rto \abo$.
We now introduce an area scale $\si_1$ in order to define a dimensionless surface tension, $\bga \equiv \bt \si_1 \ga$, and also a dimensionless parameter, $\eta = \si_1\inv \si / \Nw$, that characterizes the surface-volume ratio.
For instance, if $\si_1$ is the exposed area of a solvent molecule at the liquid-vapor interface, then $\bga$ estimates the contribution of the molecule to the surface tension and $\eta$ estimates the fraction of solvent molecules at the interface.
With these definitions, we now obtain a simple expression for the scaled surface tension, 
\begin{equation}
\label{eq-dtbga-abo}
-\dt\bga = \sumo \kbo \abo - \half \sumoop \abo \oveioop \abop + \mO(\abo^3) 
\end{equation}
where 
\begin{eqnarray}
\kbo 
	& \equiv & 
			\eta\inv \left( \rto - 1 \right) 
		= 	
			\eta\inv\left( \frac{\qto - \qbo}{\qbo} \right) 	\\
\label{def-oveioop}            
\oveioop
	& \equiv & 
			\eta\inv\left( \rto \ovetoop \rtop - \oveboop \right) 		.
\end{eqnarray}
Note that $\dt\bga$, $\kbo$, and $\oveioop$ are all rendered dimensionless by being expressed in units of $\si_1$.
It is also worth noting that the expression for $\kbo$ appealingly relates to an effective 1-solute partition function for solutes at the interface.

To this point, our treatment in Section~\ref{Sec-StatmechIntfaclProp} describes the interface between any two coexisting fluid phases.
We now specialize to the situation discussed in Section~\ref{Sec-Theory-Coexist}: a liquid solution in coexistence with a vapor phase.
The remaining task is to relate Eq.~\eqref{eq-dtbga-abo} to the properties of the liquid solution.
Section~\ref{SubSec-EffPartnFcnCoex} derives a simple approximation for doing so.
In the following, we summarize this approximate treatment.

In the bulk ensemble for pure solvent, the liquid region contains (on average) $\Nstlwz$ solvent molecules in a volume, $\Vl$, that fluctuates about an average $\Vstlz$.
For large systems, we expect that $\Vstlz$ corresponds to the average volume for $\Nstlwz$ solvent molecules at the pressure, $\Pnull$, corresponding to liquid-vapor coexistence for pure solvent, i.e.,  $\Vstlz = \Vstlz(\Pnull)$. 
We define an effective partition function, $\qlo$, describing the interaction of a single $o$-solute with solvent under these conditions: 
$\qlo 
	\equiv 
		\frac{1}{\Nstlwz} \Do(\Nstlwz,\Pnull)/\Dnull(\Nstlwz,\Pnull) 
		$,
where $\Dnull$ and $\Do$ are the isothermal-isobaric partition functions for pure solvent and for a solution with a single $o$-solute, respectively.
For sufficiently large systems, we expect that $\qlo$ should be independent of system size, i.e.,
$\qlo = \qlo(\Pnull)$.

In this bulk ensemble for pure solvent, the coexisting vapor region contains (on average) $\Nstvwz$ solvent molecules in an (average) volume of $\Vstvz$ that corresponds to the equilibrium volume for $\Nstvwz$ solvent molecules at the coexistence pressure, i.e., $\Vstvz = \Vstvz(\Nstvwz,\Pnull)$.
We define an effective partition function, $\qvo$, describing the interaction of a single $o$-solute with the vapor solvent under these conditions: 
$\qvo 
	\equiv 
		\frac{1}{\Vstvz} \Qo(\Nstvwz,\Vstvz)/\Qnull(\Nstvwz,\Vstvz) $, 
where $\Qnull$ and $\Qo$ are the canonical partition functions for pure solvent and for a solution with a single $o$-solute, respectively.
We expect that, for sufficiently large systems, $\qvo = \qvo(\rhostvwz)$, 
where $\rhostvwz = \Nstvwz/\Vstvz = \rhostvwz(\Pnull)$ is the density of pure solvent vapor at the coexistence state point.

We now relate the bulk activity of $o$-solutes to their activity in the liquid solution by defining
\begin{equation}
\label{def-alpo}
\alpo \equiv \frac{\qbo}{\qlo} = \frac{\abo}{\alo}		,
\end{equation}
where $\alo = \qlo \lao$ is the activity of $o$-solutes in a liquid solution at a constant pressure, $\Pnull$.
Section~\ref{SubSec-EffPartnFcnCoex} demonstrates that 
\begin{equation}
\label{eq-alpo-approx}
\alpo 
	= 	
		\flwnull + \fvwnull \rKo
\end{equation}
where $\flwnull = \Nstlwz/\Nw$ and $\fvwnull = \Nstvwz/\Nw$ are the fraction of solvent molecules in the liquid and vapor phases of the bulk ensemble for pure solvent.
We have also introduced the equilibrium constant
\begin{equation}
\label{def-rKo-maintext}
\rKo 
	\equiv 
		\frac{1}{\rhostvwz} \frac{\qvo}{\qlo} 
		,
\end{equation}
which  depends upon the properties of a single $o$-solute and its interactions with pure solvent in the coexisting liquid and vapor phases, but is independent of $\muo$.
We can express $\rKo$ in analogy to Henry's law:	
\begin{equation}
\label{eq-rKo-maintext}
\rKo
	= 
		\frac{\avo/\rhostvwz}{\alo}		
	= 
	\lim_{\{\la_{\op}\rightarrow0\}}
		\left[ 
		\frac{\rhovo/\rhostvwz}{\mo} 
		\right] 	
		,
\end{equation}
where $\rhovo = \Nvo / \Vv$ and $\mo = \Nlo/\Nlw$ are the density and molality, respectively, of $o$-solutes in coexisting vapor and liquid phases, while the limit indicates an infinitely dilute solution where $\la_{\op} \rightarrow 0$ for all solute species, $\op$. 
This ratio can be estimated from simulations describing coexistence between the two phases with very dilute solutes. 
While $\alpo$ reflects the partitioning of the inhomogeneous total system into liquid and vapor regions, the equilibrium constant $\rKo$ does not.

Section~\ref{SubSec-EffPartnFcnCoex} derives a simple approximation for the bulk interaction parameter,
\begin{equation}
\label{eq-oveboop-approx}
\oveboop 
	\approx 
			\alpo\inv\alpop\inv \flwnull \oveloop		,
\end{equation}
where $\oveloop$ is the thermodynamic parameter describing the interactions between infinitely dilute solutes in a closed liquid subsystem at the constant external pressure, $\Pnull$.
Here we have neglected the contribution from 
the vapor phase because we expect it should contribute negligibly when $\fvwnull \approx 0$ far from the critical point.
The factor $\alpo\inv\alpop\inv$ accounts for the tendency of infinitely dilute solutes to partition into the liquid region.
While Eqs.~\eqref{def-alpo} - \eqref{eq-oveboop-approx} are specific to liquid-vapor interfaces, we expect that the arguments of Section~\ref{SubSec-EffPartnFcnCoex} can be generalized to derive corresponding expressions for coexistence between any two coexisting fluid phases.

We can now express the surface tension variation as a perturbation expansion in terms of the composition of the coexisting liquid phase.
Specifically, Eq.~\eqref{def-alpo} allows us to relate the $\abo$ to the solute activity in the coexisting liquid phase, $\alo$.
Then Eq.~\eqref{eq-apo-mo} allows us to express $\alo$ as a perturbation expansion in the solute molality in the liquid phase, $\mo = \Nlo / \Nlw$.
This leads to 
\begin{equation}
\label{eq-dtbga-mo}
-\dt\bga 
	= 
		\sumo \ko \mo 
	- 
		\half \sumoop \mo \hoop \mop 
	+ 
		\mO(m^3) 	.
\end{equation}
Here, the intrinsic interfacial preference is
\begin{equation}
\label{def-ko}
\ko \equiv \alpo \kbo = \eta\inv \left( \frac{\qto-\qbo}{\qlo} \right)  	,
\end{equation}
while the curvature parameter is 
\begin{equation}
\label{def-hoop}
\hoop 
	\equiv 
		\alpo \oveioop \alpop - \left( \ko + \kop \right) \oveloop 	.
\end{equation}
These two results are also physically sensible.
Eq.~\eqref{def-ko} expresses $\ko$ as the ratio of the effective one-solute partition function for the interface, $\qio \equiv \qto - \qbo$, with the corresponding effective partition function for the coexisting liquid phase, $\qlo$.
Similarly, the first term in Eq.~\eqref{def-hoop} scales the interfacial interaction energy, $\oveioop$, by the tendency of the solutes to partition into the liquid phase, while the second term reflects bulk solute interactions and their tendency to partition to the interface.
Importantly, Eq.~\eqref{eq-dtbga-mo} is entirely analogous to the result that we obtained earlier for a simpler lattice model that did not account for  two-phase coexistence.

\newcommand{\bGa}{\overline \Ga}
\newcommand{\bGao}{\bGa_o}
\newcommand{\bGaop}{\bGa_\op}
\newcommand{\setmo}{\{\mo\}}
\newcommand{\notmo}{m_{\hat o}}
\newcommand{\notmop}{m_{\hat o'}}
\newcommand{\mK}{{\mathbb{K}}}
\newcommand{\mKoop}{\mK_{\oop}}
\newcommand{\rb}{{\rm b}}

We now seek to obtain a simple equation for the surface excess of $o$-solutes.
The Gibbs adsorption equation, Eq.~\eqref{eq-Gibbs-adsorption-eqn}, may be re-expressed (at constant temperature):
\begin{equation}
\label{eq-scaled-ddtbga}
\dd\dt\bga 
	=
	- 
		\sumo \bGao \dd\bmuo 	,
\end{equation}
where we have defined a scaled surface excess, $\bGao \equiv \si_1 \Gao = \frac{\si_1}{\si} \left( \Nto - \Nbo \right)$, and a scaled chemical potential, $\bmuo = \bt \muo$. 
From Eq.~\eqref{eq-dtbga-mo}, we have
\begin{equation}
\label{def-bo}
\rb_o 
	\equiv 
			\left( 
				\frac{\ptl\bga}{\ptl\mo}
			\right)_{T,\notmo}
	= 
		-
			\ko 
		+ 
			\sumop 
					\hoop \mop
		+ 
			\mO(m^2) 	,
\end{equation}
where the subscript $\notmo$ indicates that the derivative is evaluated while holding $m_{\op}$ constant for all $\op \neq o$.
We now approximate the chemical potential, $\bmuo(T,\setmo)$, of the coexisting liquid phase by the chemical potential, $\bmuo(T,\Pnull,\setmo)$, of the bulk liquid phase at the fixed pressure, $\Pnull$.
Following the arguments of Section~\ref{SubSec-ApproxChemPotlDeriv}, we use Eq.~\eqref{eq-btmuo-bulk-liquid} to evaluate the derivative
\begin{equation}
\label{def-mKoop}
\mKoop 
	\equiv
		\left( 
			\frac{\ptl \bmuo}{\ptl\mop}
		\right)_{T,\Pnull,\notmop} 
	= 
		m\inv_o \de_{\oop}
		+
		\oveloop  
		+ 
		\mO(m) 
	= 
		\mK_{\op o}	.
\end{equation}
Using this approximation, Eqs.~\eqref{eq-scaled-ddtbga}-\eqref{def-mKoop} imply that 
\begin{equation}
\label{eq-bo}
\rb_o 
	\approx
		- 
		\sumop \mKoop	 \bGa_\op		,
\end{equation}
for variations at constant temperature.
We invert this system of equations to determine the adsorption coefficients:
\begin{equation}
\bGa 
	\approx
		- \mK\inv \rb		.
\end{equation}
For systems with a single type of $o$-solute, we obtain at lowest order
\begin{equation}
\label{eq-bGa-onesolute}
\bGa 
	\approx
		m
		\left( 
			\frac{k-hm}{1+\ove m}
		\right)
		,
\end{equation}
where we have suppressed the subscript $o$.
For systems with two types of $o$-solutes, we similarly obtain
\begin{equation}
\label{eq-bGa-twosolutes}
\bGa_1 
	\approx
		m_1
		\left[ 
			\frac{
				k_1 
				- h_{11} m_1 
				- \left( h_{12} - k_1 \ove_{22} + k_2 \ove_{12} \right) m_2
				}
				{
				1 + \ove_{11} m_1 + \ove_{22} m_2 
				}
		\right]
\end{equation}
and an analogous equation for $\bGa_2$.

\newcommand{\rhow}{\rho_w}
\newcommand{\rhowphist}{\rho_{w\phi}^*}
\newcommand{\Qop}{Q_{o'}}
\newcommand{\Qoop}{Q_{oo'}}
\newcommand{\qvoinfty}{q_{o}^{\infty}}
\newcommand{\evoop}{\varepsilon_{v;\oop}}
\newcommand{\beoop}{\overline\varepsilon_{\oop}}
\newcommand{\bevoop}{\overline\varepsilon_{v;\oop}}
\newcommand{\tevoop}{\tilde{\varepsilon}_{v;\oop}}
\newcommand{\tevoopinfty}{\tevoop^{\;\infty}}
\newcommand{\rhoo}{\rho_o}
\newcommand{\rhoop}{\rho_{o'}}

\newcommand{\Boop}{B_{\oop}}
\newcommand{\Boopinfty}{\Boop^{\infty}}
\newcommand{\Bvoop}{B_{v;\oop}}
\newcommand{\Bvoopinfty}{\Bvoop^{\;\infty}}

\newcommand{\qvolo}{q_o}
\newcommand{\qvolop}{q_{o'}}
\newcommand{\avolo}{a_o}

\subsection{Effective 1- and 2-solute partition functions}
\label{SubSec-EffPartnFcnCoex}
\subsubsection{Dilute solutions at constant volume}
We consider a one-phase system with $\Nw$  solvent molecules in a volume, $V$, at a  temperature, $T$.
We denote the canonical partition function for the pure solvent system by $\Qnull = \Qnull(\Nw,V,T)$.
We define an effective (scaled) one $o$-solute partition function
\begin{equation}
\label{eq-qvo}
\qvolo = \qvolo(\Nw,V,T) = \frac{1}{V} \frac{\Qo}{\Qnull}	,
\end{equation}
where $\Qo = \Qo(\Nw,V,T)$ is the canonical partition function when a single $o$-solute is added to the system.
We define the solute activity 
\begin{equation}
\label{def-avolo}
\avolo = \qvolo \lao	,
\end{equation}
such that $\qvolo$ and $\avolo$ have units of inverse volume. 
We define $\Qoop = \Qoop(\Nw,V,T)$ as the canonical partition when the system includes exactly two solutes, one of which is an $o$-solute and the second is an $o'$-solute. 
The parameter
\begin{equation}
\label{def-Bvoop} 	
\Boop
	= 
		 \frac{1}{V} 
			\left[ 
				\Coop \frac{\Qoop}{\Qnull} 
			- 
				\frac{\Qo}{\Qnull}\frac{\Qop}{\Qnull}
			\right]
\end{equation}
corresponds to an effective partition function describing the solvent-mediated interaction between the two solutes.
In the thermodynamic limit, we expect that $\qvolo$ and $\Boop$ depend only upon the solvent density, $\rhow=\Nw/V$, and temperature: $\qvolo \rightarrow \qvolo^{\;\infty}(\rhow,T)$ and  $\Boop \rightarrow \Boopinfty(\rhow,T)$.
The activity of $o$-solutes may then be expressed as a power series in the solute density, $\rhoo = N_o/V$:
\begin{equation}
\label{eq-avo-rho}
\avolo 
	= 
		\rhoo 
		\left[ 	
			1 + \sumop \beoop \rho_\op + \mO(\rhoo^2)
		\right]
		,
\end{equation}
where 
we have defined the effective solute-solute interaction parameter 
\begin{equation}
\label{def-beoop}
\beoop = - \qvolo\inv \qvolop\inv \Boop 			.
\end{equation}
Equation~\eqref{def-beoop} adopts the convention that $\beoop < 0$ for attractive interactions. 
The chemical potential of $o$-solutes is then
\begin{equation}
\bt\muo = \ln \left[ \rhoo/\qvolo \right] + \sumop \beoop \rhoop + \mO(\rhoo^2)		.
\end{equation}

\subsubsection{Dilute solutions at constant pressure}
\label{subsec-App-ConstPressSoln}
We now consider a corresponding one-phase system with $\Nw$ solvent molecules at the external pressure, $P$, and  temperature, $T$.
The partition function for pure solvent in this isothermal-isobaric ensemble is $\Dnull \equiv \Dnull(\Nw,P,T)$.
The corresponding partition functions for solutions with the addition of one and two cosolutes are $\Do \equiv \Do(\Nw,P,T)$ and $\Doop \equiv \Doop(\Nw,P,T)$, respectively.
We define the effective (scaled) one $o$-solute partition function and the solute activity:
\begin{eqnarray}
\label{def-qpo}
\qpo 	& = & \frac{1}{\Nw} \frac{\Do}{\Dnull}		\\
\label{def-apo}
\apo	& = & \qpo \lao	.
\end{eqnarray}
We use the symbols $\qpo$ and $\apo$ to emphasize the analogy with Eqs.~\eqref{eq-qvo} and \eqref{def-avolo} for the canonical ensemble, while using the subscript `$p$' to indicate that these quantities correspond to the constant pressure ensemble. 
The effective two-solute partition function for the constant pressure ensemble is
\begin{equation}
\label{def-Bpoop} 
\Bpoop
	= 
		 \frac{1}{\Nw} 
			\left[ 
				\Coop \frac{\Doop}{\Dnull} 
			- 
				\frac{\Do}{\Dnull}\frac{\Dop}{\Dnull}
			\right]		.
\end{equation}
In the thermodynamic limit, we expect that $\qpo$ and $\Bpoop$ depend only upon the pressure and temperature: $\qpo \rightarrow \qpo^{\infty}(P,T)$ and  $\Bpoop \rightarrow \Bpoopinfty(P,T)$.
The activity of $o$-solutes may be expressed in analogy to Eq.~\eqref{eq-avo-rho}:
\begin{equation}
\label{eq-apo-mo}
\apo
	= 
		\mo 
		\left[ 
			1 + \sumop \bepoop \mop + \mO(\mo^2)
		\right]
		,
\end{equation}
where $\mo = N_o/\Nw$ is the molality of $o$-solutes and we have defined
\begin{equation}
\label{def-App-bepoop}
\bepoop = - \qpo\inv \qpop\inv \Bpoop 				.
\end{equation}
The chemical potential of $o$-solutes is then 
\begin{equation}
\label{eq-btmuo-bulk-liquid}
\bt\muo = \ln \left[ \mo /\qpo \right] + \sumop \bepoop \mop + \mO(m_o^2)		.
\end{equation}

\newcommand{\bbP}{\mathbb{P}}
\newcommand{\bbPp}{\bbP_{p\phi}}
\newcommand{\Vphist}{V_\phi^*}

We now relate the constant volume and constant pressure ensembles at fixed $\Nw$ and $T$.
We first relate $\qvolo$ and $\qpo$.
We define $\bbPp(V)$ as the probability that the pure solvent system samples the volume, $V$ at the given pressure, $P$:
\begin{equation}
\label{def-bbPp}
\bbPp(V) = \bbPp(V;\Nw,P,T) = \left. \Qnull(V) e^{-\bt P V} \right/ \Dnull(P)	.
\end{equation}
We define the ratio of canonical partition functions with and without an $o$-solute:
\begin{equation}
\label{def-ro}
\ro(V) \equiv \ro(\Nw,V,T) = \Qo(\Nw,V,T) / \Qnull(\Nw,V,T)		,
\end{equation}
such that  $\qvolo(V) =  \ro(V) / V$.
Using standard approaches from thermodynamic perturbation theory, we may express
\begin{equation}
\label{eq-qpo}
\qpo = \frac{1}{\Nw} \llangle \ro(V) \rrangle_{p\phi}		,
\end{equation}
where the average is evaluated according to $\bbPp(V)$.
In the thermodynamic limit, we expect that $\bbPp(V) \rightarrow \de(V-\Vphist)$, where $\Vphist = \Vphist(\Nw,P,T)$ is the equilibrium thermodynamic volume of pure solvent at the given $(\Nw,P,T)$.
Thus, in this limit, we expect that 
\begin{equation}
\label{eq-qpo-qvoinfty}
\qpo \rightarrow \qvoinfty(\rhowphist,T) / \rhowphist		,
\end{equation}
where $\rhowphist = \rhowphist(P,T) = \Nw/\Vphist$ is the thermodynamic density of pure solvent at the given pressure and temperature.

\newcommand{\roop}{r_{\oop}}
\newcommand{\chipoop}{\chi^2_{p;\oop}}
\newcommand{\sioop}{\si^2_{p;\oop}}
\newcommand{\brpo}{\overline r_{p;o}}
\newcommand{\brpop}{\overline r_{p;o'}}
\newcommand{\dro}{\de r_{o}}
\newcommand{\drop}{\de r_{\op}}
\newcommand{\drpo}{\de r_{p;o}}
\newcommand{\drpop}{\de r_{p;o'}}

We now relate $\Boop$ and $\Bpoop$.
We define a corresponding ratio for two solutes:
\begin{equation}
\label{def-roop}
\roop(V) \equiv \roop(\Nw,V,T) \equiv \Qoop(\Nw,V,T) / \Qnull(\Nw,V,T)	. 
\end{equation}
We can then express the two-solute contribution to $\Bpoop$ by $\Doop / \Dnull = \llangle \roop(V) \rrangle_{p\phi}$.
However, eliminating the one-solute contributions is more subtle and should account for correlated fluctuations.
Accordingly, we define 
\begin{equation}
\label{def-chipoop}
\chipoop 
	\equiv 
		\chipoop(\Nw,P,T) 
	\equiv 
		\llangle \dro(V) \drop(V) \rrangle_{p\phi}	
	= 
		\llangle \ro(V) \rop(V) \rrangle_{p\phi} 
	- 
		\llangle \ro(V) \rrangle_{p\phi}
		\llangle \rop(V) \rrangle_{p\phi}
\end{equation}
where 
$\dro(V) = \ro(V) -  \llangle \ro(V) \rrangle_{p\phi}$.
We recognize the last term in Eq.~\eqref{def-chipoop} as the product $\left( \Do/\Dnull \right) \times \left( \Dop / \Dnull \right)$ in Eq.~\eqref{def-Bpoop}. 
Consequently, using Eq.~\eqref{def-chipoop} to eliminate this product, we obtain
\begin{equation}
\Bpoop(\Nw,P,T) 
	= 
		\frac{1}{\Nw}
		 \llangle V \Boop(V) \rrangle_{p\phi}
	+ 
		\frac{1}{\Nw}
		\chipoop(\Nw,P,T)		.
\end{equation}
In the thermodynamic limit, we expect that $\Bpoop(\Nw,P,T)$ should converge upon 
\begin{equation}
\Bpoopinfty(P,T) 
	= 
		\frac{1}{\rhowphist}
		\Boopinfty(\rhowphist,T)
	+ 	\sioop(P,T)
\end{equation}
where  
\begin{equation}
\sioop(P,T) = \lim_{\Nw\rightarrow\infty} \Nw\inv \chipoop(\Nw,P,T)	,
\end{equation}	
such that both $\Bpoopinfty$ and $\sioop$ are intensive.

\subsubsection{Dilute bulk solutions}
\label{subsec-App-BulkSolns}
We now consider the bulk ensemble describing coexistence between liquid and vapor subsystems.
The key thermodynamic parameters for this bulk ensemble are $\qbo$ and $\oveboop$, which are defined by Eqs.~\eqref{def-qbo} and \eqref{def-oveboop}, respectively. 
Here we follow the argument in Section~\ref{subsec-App-ConstPressSoln} to relate $\qbo$ and $\oveboop$ to corresponding parameters for the coexisting phases.

\newcommand{\mPbnull}{\mP_{b\phi}}
We first consider the bulk system in the absence of solutes.
In this pure solvent case, the probability that the liquid region contains $\Nlw$ solvent molecules in a volume $\Vl$ is given by
\begin{equation}
\label{def-mPbnull}
\mPbnull(\Nlw,\Vl) = \Qlnull(\Nlw,\Vl) \Qvnull(\oNvw,\oVv) / \Qbnull
\end{equation}
where $\oNvw = \Nw - \Nlw$ and $\oVv = V - \Vl$.
For large systems, we expect that $\mPbnull(\Nlw,\Vl) \rightarrow \dt( \Nlw - \Nstlwz ) \dt( \Vl - \Vstlz)$, where $\Nstlwz$ and $\Vstlz$ are the thermodynamic values for $\Nlw$ and $\Vl$ in the bulk ensemble for pure solvent.  
Moreover, we expect that $\Vstlz$ corresponds to the equilibrium volume of $\Nstlwz$ solvent molecules in the constant pressure ensemble at the pressure $P = \Pnull$ corresponding to liquid-vapor coexistence for pure solvent, i.e., $\Vstlz = \Vstlz(\Nstlwz,\Pnull,T)$.

\newcommand{\rlo}{r_\lo}
\newcommand{\rlop}{r_\lop}
\newcommand{\rvo}{r_\vo}
\newcommand{\rvop}{r_\vop}
\newcommand{\rbo}{r_\bo}
\newcommand{\rbop}{r_{b\op}}
\newcommand{\drlo}{\de\rlo}
\newcommand{\drlop}{\de\rlop}
\newcommand{\drvo}{\de\rvo}
\newcommand{\drvop}{\de\rvop}
\newcommand{\drbo}{\de\rbo}
\newcommand{\drbop}{\de\rbop}

In order to treat the ratio $\Qbo/\Qbnull$, we first define 
\begin{eqnarray}
r_\lo 		
	& = & 
		r_\lo(\Nlw,\Vl) = Q_\lo(\Nlw,\Vl)/\Qlnull(\Nlw,\Vl)		
\\
r_\vo 	
	& = & 
		r_\vo(\oNvw,\oVv) = Q_\vo(\oNvw,\oVv)/\Qvnull(\oNvw,\oVv)	
,
\end{eqnarray} 
which correspond to the canonical partition function ratio $\ro$ in Eq.~\eqref{def-ro} for the liquid and vapor regions, respectively. 
We also define 
\begin{equation}
\rbo 
	= 
		\rbo(\Nlw,\Vl) 
	= 
		\rlo(\Nlw,\Vl) 
	+ 
		\rvo(\oNvw,\oVv)		
		. 	
\end{equation}
We then have
\begin{equation}
\label{eq-QboQbnull-ratio}
\frac{\Qbo}{\Qbnull} 
	= 
		\llangle
				\frac{Q_\lo}{\Qlnull} 
			+	
				\frac{Q_\vo}{\Qvnull} 		
		\rrangle_{b\phi}			
	=
		\llangle 
			\rlo
		\rrangle_{b\phi}			
	+
		\llangle
			\rvo
		\rrangle_{b\phi}		
	= 
		\llangle 
			\rbo 	
		\rrangle_{b\phi}		
	,
\end{equation}
where the averages are evaluated over the bulk ensemble for pure solvent according to Eq.~\eqref{def-mPbnull}.
We expect that for large systems we may approximate
\begin{eqnarray}
\label{eq-rlo-bphi}
\llangle \rlo \rrangle_{b\phi}
	& = & 
		\llangle \rlo(\Nstlwz,\Vl) \rrangle_{\Nstlwz \Pnull T}		\\
\label{eq-rvo-bphi}
\llangle \rvo \rrangle_{b\phi}
	& = &
		\llangle \rvo(\Nstvwz,\Vv) \rrangle_{\Nstvwz \Pnull T}	
	=		\rvo(\Nstvwz,\Vstvz) 	.
\end{eqnarray}
The angular brackets in Eqs.~\eqref{eq-rlo-bphi} and \eqref{eq-rvo-bphi} denote isothermal-isobaric ensemble averages at $(\Pnull,T)$ for  $\Nstlwz$ solvent molecules in the liquid phase and for $\Nstvwz$ solvent molecules in the vapor phase, respectively. 
We expect that for large systems
\begin{equation}
\frac{\Qbo}{\Qbnull} 
	= 
		\Nstlwz \qlo + \Vstvz \qvo
\end{equation}
where $\Vstvz = \Vstvz(\Nstvwz,\Pnull,T) =   V - \Vstlz$ is the equilibrium volume of the vapor region.
Here $\qlo = q_{l; p o }^{\infty}(\Pnull,T)$ is the thermodynamic limit of the constant pressure scaled partition function, $\qpo$, for one $o$-solute in the liquid phase at the state point $(\Pnull,T)$ for liquid-vapor coexistence of pure solvent. 
Similarly, $\qvo = \qvoinfty(\rhostvwz,T)$ is the thermodynamic limit of the constant volume scaled partition function, $\qvolo$, for one $o$-solute in the vapor phase at the solvent density, $\rhostvwz = \rhostvwz(\Pnull,T)$, corresponding to liquid-vapor coexistence for pure solvent.
We define the equilibrium constant 
\begin{equation}
\label{def-rKo}
\rKo \equiv \frac{1}{\rhostvwz} \frac{\qvo}{\qlo}		
.
\end{equation}
$\rKo$ is analogous to the Henry's law constant that describes the partitioning of infinitely dilute solutes between coexisting vapor and liquid phases of pure solvent.
This allows us to relate the one-solute effective partition function, $\qbo$, for the bulk ensemble to the corresponding effective partition function, $\qlo$, for the liquid phase at constant pressure:
\begin{equation}
\label{def-app-alpo}
\alpo 
	\equiv 
		\frac{\qbo}{\qlo} 
	=
		\flwnull + \fvwnull \rKo 
\end{equation}
where 
$\flwnull = \Nstlwz/\Nw$ and $\fvwnull = \Nstvwz/\Nw$ are the fraction of solvent molecules in the liquid and vapor phases of the pure solvent bulk ensemble.
Note that $\rKo = \rKo(\Pnull,T)$ depends only upon the interactions of a single $o$-solute with solvent and the intensive properties determining liquid-vapor coexistence for pure solvent.
In contrast, $\alpo$ also depends upon the equilibrium partitioning of the pure solvent bulk ensemble between the liquid and vapor phases. 

In order to treat the bulk interaction parameter, we define the corresponding effective partition function for two solutes:
\begin{equation}
\label{def-Bboop} 	
\Bboop
	= 
		 \frac{1}{\Nw} 
			\left[ 
				\Coop \frac{\Qboop}{\Qbnull} 
			- 
				\frac{\Qbo}{\Qbnull}\frac{\Qbop}{\Qbnull}
			\right]		,
\end{equation}
such that $\oveboop = - \qbo\inv \qbop\inv \Bboop $.
The first term in Eq.~\eqref{def-Bboop} may be decomposed into different partitions of the solutes between the two phases:
\begin{equation}
\Coop \frac{\Qboop}{\Qbnull}
	= 
		\llangle 
			\left[
				\Coop \rloop 
				+ 
				\rlo \rvop
				+ 
				\rlop \rvo
				+
				\Coop \rvoop
			\right]
		\rrangle_{b\phi}	,
\end{equation}
where 
$\rloop = \rloop(\Nlw,\Vl) = \Qloop(\Nlw,\Vl)/\Qlnull(\Nlw,\Vl)$ 
and 
$\rvoop = \rvoop(\oNvw,\oVv) = \Qvoop(\oNvw,\oVv)/\Qvnull(\oNvw,\oVv)$ 
are partition function ratios  for adding both solutes to the liquid and vapor phases, respectively.
We now treat bulk ensemble fluctuations in analogy to Eq.~\eqref{def-chipoop} for the constant pressure ensemble: 
\begin{equation}
\label{def-chiboop}
\chiboop 
	\equiv 
		\llangle \drbo \drbop \rrangle_{b\phi}	
	= 
		\llangle \rbo \rbop \rrangle_{b\phi} 
	- 
		\frac{\Qbo}{\Qbnull}
		\frac{\Qbop}{\Qbnull}
\end{equation}
where 
$\drbo \equiv \drbo(\Nlw,\Vl) \equiv \rbo(\Nlw,\Vl) -  \llangle \rbo(\Nlw,\Vl) \rrangle_{b\phi}
= 
\drlo + \drvo $
with 
$\drlo = \rlo - \llangle \rlo \rrangle_{b\phi}$ 
and 
$\drvo = \rvo - \llangle \rvo \rrangle_{b\phi}$.
For large systems, we expect that averages over the bulk ensemble may be approximated by averages over the corresponding isothermal-isobaric ensemble:
\begin{equation}
\label{eq-chiboop}
\chiboop 
	\approx		
			\chiloop(\Nstlwz,\Pnull) 
		+ 
			\chivoop(\Nstvwz,\Pnull)
		+ 
			\chixoop	,
\end{equation}
where 
$\chiloop(\Nstlwz,\Pnull) = \llangle \drlo \drlop \rrangle_{\Nstlwz \Pnull T}$ and 
$\chivoop(\Nstvwz,\Pnull) = \llangle \drvo \drvop \rrangle_{\Nstvwz \Pnull T}$ describe fluctuations in the liquid and vapor phases, respectively, of pure solvent.
The third term in Eq.~\eqref{eq-chiboop}
\begin{equation}
\chixoop 
	\equiv 		
			\llangle \drlo \drvop + \drlop \drvo \rrangle_{b\phi}	
	\approx 
			0 
\end{equation} 
describes correlated fluctuations between the liquid and vapor regions of the bulk ensemble.
We assume $\chixoop$ should vanish in the thermodynamic limit because both regions are assumed far from the interface.
Using Eq.~\eqref{def-chiboop} to eliminate the product 
$\frac{\Qbo}{\Qbnull} \times \frac{\Qbop}{\Qbnull}$ from Eq.~\eqref{def-Bboop}, we obtain
\begin{eqnarray}
\nonumber
\Nw \Bboop 
	& \approx &  
			\llangle 
				\Coop \rloop - \rlo \rlop
			\rrangle_{b\phi} 	
			+ 
			\chiloop(\Nstlwz,\Pnull) 			\\
	& +  & 
			\llangle 
				\Coop \rvoop - \rvo \rvop
			\rrangle_{b\phi} 	
			+ 
			\chivoop(\Nstvwz,\Pnull) 			.			
\end{eqnarray}
For large systems, we expect that the first and second lines should reduce to $\Nstlwz \Bloop(\Nstlwz,\Pnull)$ and $\Nstvwz \Bvoop(\Nstvwz,\Pnull)$, respectively.
Thus, the interaction parameter for the bulk ensemble may be approximated
\begin{equation}
\label{eq-App-oveboop}
\oveboop
	\approx
		\alpo\inv \alpop\inv
		\left[ 
			\flwnull \ovelloop(\Pnull)
		+ 
			\fvwnull 			\Ko \Kop \ovevoop(\Pnull)
		\right]
.
\end{equation}
Here 
$\ovelloop(\Pnull)$ and $\ovevoop(\Pnull)$ are interaction parameters defined according to Eq.~\eqref{def-App-bepoop} for the liquid and vapor phases at the state point $(\Pnull,T)$ corresponding to liquid-vapor coexistence for pure solvent.
In the main text, we neglect the second term describing interactions in the vapor phase because $\fvwnull \approx 0$.

\subsection{Approximating derivatives at coexistence}
\label{SubSec-ApproxChemPotlDeriv}
First consider a homogeneous solution of solvent ($w$) with $M$ cosolute species, $o = 1, \ldots, M$. 
The equilibrium state is specified by $M+3$ variables, e.g., $(\Nlw, \setNlo,V,T)$.
The intensive state is specified by $M+2$ variables, e.g., $(T,P,\setmo)$.
In particular, the chemical potential of each species $i = w$ or $o$ is a function $\mu_i = \mu_i(T,P,\setmo)$.
The total differential is 
\begin{equation}
\dd \mu_i = \dd \mu_i(T,P,\setmo) = - \oS_i \dd T + \oV_i \dd P + \sumo \omu_{io} \dd \mo	,
\end{equation} 
where $\oS_i$ is the partial molar entropy, $\oV_i = (\ptl V/ \ptl n_i)_{T,P,n_{j\neq i}}$ is the partial molar volume, and $\omu_{io} = (\ptl \mu_i / \ptl \mo)_{T,P,\{m_{o'\neq o}\}}$.

Now consider an interfacial system with coexistence between liquid and vapor phases in a constant volume, $V$.
The equilibrium state is then specified by, e.g., $(\Nw,\setNto,V,T,\si)$.
The coexistence criteria are 
\begin{equation}
\label{eq-coexist-two}
\begin{array}{lcl}
P_l(\rholw,\setmo,T) 		& = & P_v(\rhovw,\setrhovo,T)		= P 		\\
\mulw(\rholw,\setmo,T) 	& = & \muvw(\rhovw,\setrhovo,T)		= \muw	\\
\mulo(\rholw,\setmo,T) 	& = & \muvo(\rhovw,\setrhovo,T)		= \muo  \hspace{0.5in}
											\text{for $o = 1,\ldots, M$},
\end{array}
\end{equation}
This is a system of $M+2$ equations for $2(M+1)+1 = 2M+3$ variables.
Consequently, we expect an $M+1$ dimensional family of solutions that are uniquely determined by $(\setmo,T)$.
Thus, for any $(\setmo,T)$, there is a unique equilibrium state with coexistence between liquid and vapor phases.
This corresponds to the Gibbs phase rule: a system with $c = M+1$ components and $p = 2$ has $ f = c - p +2 = M + 1 - 2 + 2 = M + 1$ independent thermodynamic degrees of freedom.
As in the main text, we shall assume $T$ is constant and (usually) treat it implicitly.

Since we have defined this equilibrium state to have constant $V$, the pressure $P = \Plv(\setmo)$ varies with the composition. 
Consequently, 
\begin{equation}
\mu_{lv; w} = \mulw(P,\setmo) = \mulw\left(\Plv(\setmo),\setmo\right) = \mu_{lv;w}(\setmo)
.
\end{equation}
Similarly, $\mu_{lv;o}$ is uniquely specified by $\setmo$.

We now wish to consider the surface tension, $\ga$.
The Gibbs adsorption equation at constant $T$ is 
\begin{equation}
\label{eq-Gibbs-adsorption-eqn}
\dd\ga = - \sumo \Gao \dd\muo	.
\end{equation}
Note that $\ga$ is a function of $M+1$ variables, $(T,\setmo)$ and we must understand $\muo = \muo(T,\setmo) = \muo(T, \Plv(T,\setmo),\setmo)$.

In the main text, we evaluate
\begin{eqnarray}
\left( \frac{\ptl \bga}{\ptl \mo} \right)_{T,\{m_{o'\neq o}\}}
 	& = &
		- \sumop 
				\bGaop 
				\left( \frac{\ptl \bmu_\op}{\ptl \mo} \right)_{T,\{m_{o''\neq o}\}}
	\\
	& \approx & 
		- \sumop 
				\bGaop 
				\left( \frac{\ptl \bmu_\op}{\ptl \mo} \right)_{T,P,\{m_{o''\neq o}\}}	,
\end{eqnarray}
where in the last line we make the approximation of evaluating the derivative at constant pressure, $P$. 
However, this is clearly an approximation based upon the preceding, i.e., $P = \Plv(T,\setmo)$ varies with $\mo$ at coexistence.

The preceding discussion also makes it clear how to address this.
At constant temperature,
\begin{equation}
\bmuo(\setmo) = \bmuo\left(\Plv(\setmo),\setmo\right)		.
\end{equation}
Consequently, for coexistence at constant temperature, we have
\begin{equation}
\label{eq-dbmuo}
\dd\bmuo = \bt \oV_{lo} \dd \Plv + \sumop \bt \omu_{\oop} \dd\mop		.
\end{equation}
We now have to determine how $\Plv$ varies with $\setmo$.

We can make a simple calculation for $\dd\Plv$, if we assume that the vapor phase is ideal.
In this case, we have
\begin{equation}
\bt \Plv = \rhovw + \sumo \rhovo 	.
\end{equation}
If we consider very dilute solutions, we can use our Henry's law-type constant, $\rKo$, to express $\rhovo = \rhostvwz \rKo\mo$.
Thus, at constant temperature,
\begin{equation}
\label{eq-dPlv-1}
\dd \left( \bt \Plv \right) = \dd \rhovw + \rhostvwz \sumo \rKo \dd\mo 	.
\end{equation}
Moreover, following the preceding assumptions, we have
\begin{equation}
\bmuw = \ln [ \rhovw / q_{vw} ] = \bmu_{lw}(T,\Plv) - \sumo \mo	,
\end{equation}
where the middle expression corresponds to an ideal gas, while the right expression corresponds to an infinitely dilute solution.
Consequently, (at constant $T$) we have
\begin{equation}
\label{eq-dPlv-2}
\rhovw\inv \dd \rhovw 	= \bt \oV_{lw} \dd \Plv - \sumo \dd \mo
\end{equation}
(If we were interested in dilute interacting gases, then we would have to include second order terms in the virial expansion and a corresponding term in the chemical potential for solvent vapor.)

Combining Eqs.~\eqref{eq-dPlv-1} and \eqref{eq-dPlv-2}, while also expanding about the reference solution so that $\rhovw \rightarrow \rhostvwz$, we have
\begin{equation}
\label{eq-dPlv-3}
\dd\left( \bt \Plv \right) 
	= 
		\frac{1}{\de \oV_w} \sumo \left( \rKo - 1 \right) \dd\mo	,
\end{equation}
where $\de\oV_w = \frac{1}{\rhostvwz} - \oV_{lw}$, which is the difference in the partial molar volume of water in the vapor and liquid phases. 
We expect that $\de\oV_w \approx 1/ \rhostvwz$.

Using Eq.~\eqref{eq-dPlv-3} in Eq.~\eqref{eq-dbmuo}, we obtain
\begin{equation}
\dd\bmuo = \sumop \omK_\oop \dd \mop
\end{equation}
where 
\begin{equation}
\label{def-omKoop}
\omK_\oop = \mK_\oop + \frac{\oV_{lo}}{\de \oV_w}\left( \rK_\op - 1 \right)		.
\end{equation}
In contrast to $\mK_\oop$, $\omK_\oop$ is {\bf not} symmetric.

In Eq.~\eqref{def-omKoop} $\oV_{lo}$ is the partial molar volume of $o$-solutes in the liquid phase, while $\de \oV_w$ is approximately the partial molar volume of solvent in the vapor phase.
Near the triple point for water we expect that $\oV_{lo}/\de \oV_w \approx 0.01$ or smaller.
Consequently, we expect that our constant pressure approximation should be quite reasonable near the triple point
\begin{equation}
\omK_\oop \approx \mK_\oop	.
\end{equation}

In the future it may be useful to consider coexistence at constant $T$ and $P$. 
In this case, the $M$ composition variables, $\{m_1, m_2, \ldots, m_M\}$, are no longer  independent. 
Conversely, if we have a macromolecule M along with $M$ cosolutes, then $c = M+2$ and we have $f = M+2$ independent thermodynamic degrees of freedom, which we can take to be $T, P$ and the concentration of the $M$ cosolutes, $\setmo$.

\section{Details of Free Energy Perturbation Calculations}
\subsubsection{Vapor phase}
\label{subsubsec-FreeEnergyMethods-Vapor}
\newcommand{\La}{\Lambda}
\newcommand{\Lao}{\La_o}
\newcommand{\Law}{\La_w}
\newcommand{\hZ}{\widehat Z}
\newcommand{\Yvo}{Y_{vo}}
\newcommand{\Ylo}{Y_{lo}}
\newcommand{\Yto}{Y_{to}}
\newcommand{\Yloop}{Y_{l\oop}}
\newcommand{\Ytoop}{Y_{t\oop}}
\newcommand{\Pext}{P_{\rm ext}}

\newcommand{\br}{{\bf\rm r}}

\newcommand{\tVl}{\tilde V_l}
\newcommand{\tVv}{\tilde V_v}
\newcommand{\tVi}{\tilde V_i}
\newcommand{\tVb}{\tilde V_b}
\newcommand{\tNi}{\tilde N_i}
\newcommand{\tNl}{\tilde N_l}
\newcommand{\tNv}{\tilde N_v}
\newcommand{\tNb}{\tilde N_b}

\newcommand{\rhownull}{\rho_{w\phi}}
\newcommand{\Drhownull}{\De \rhownull}

The effective partition function, $\qvo$, for adding one $o$-solute into the vapor phase is
\begin{equation}
\label{def-qvo-methods}
\qvo    
    \equiv 
            \frac{1}{\Vstvz} \frac{\Qo(\Nstvwz,\Vstvz)}{\Qnull(\Nstvwz,\Vstvz)}
    =
            \frac{1}{\Vstvz} 
                \frac{\Qo(\Nstvwz,\Vstvz)}{\Qnull(\Nstvwz+1,\Vstvz)}
                    e^{-\bt \muwnull}   
            ,
\end{equation}
where $\Nstvwz$ is the number of solvent molecules in the vapor region of the bulk ensemble for pure solvent, while $\Vstvz$ is the volume of this vapor region.
We obtain the last expression by introducing $1 = \Qnull(\Nstvwz+1,\Vstvz)/\Qnull(\Nstvwz+1,\Vstvz)$ and recognizing $\Qnull(\Nstvwz+1,\Vstvz) / \Qnull(\Nstvwz,\Vstvz) = e^{-\bt \muwnull}$, where $\muwnull$ is the chemical potential of pure solvent at liquid-vapor coexistence.
The remaining ratio of partition functions may be re-expressed
\begin{equation}
\label{eq-Qo-ratio-methods}
\frac{\Qo(\Nstvwz,\Vstvz)}{\Qnull(\Nstvwz+1,\Vstvz)}
    =
        \left(\Nstvwz+1\right) 
            \left( \frac{\Law}{\Lao}\right)^3 
                \Yvo(\Nstvwz+1,\Vstvz)    ,
\end{equation}
where $\Law$ and $\Lao$ are the thermal de Broglie wavelengths for solvent and $o$-solute molecules, respectively, while $Y_{vo}(N+1,V)$ is defined by the ratio of canonical configuration integrals, $\hZ_{vo}(N,V)/\hZ_{v\phi}(N+1,V)$, associated with converting one solvent molecule into an $o$-solute for a system with $N+1$ solvent molecules in a volume $V$ :
\begin{eqnarray}
\label{def-Yvo}
\Yvo(N+1,V)   
    & \equiv &  
        \frac{\hZ_{vo}(N,V)}{\hZ_{v\phi}(N+1,V)}
    =
      \llangle 
        \frac{1}{N+1}
        \sum_{i=1}^{N+1}
            \exp\left[
                    -\bt \De U_{o|i}(\br^{N+1})
                    \right]
      \rrangle_{N+1,V,\phi}       .
\end{eqnarray}
The angular brackets in Eq.~\eqref{def-Yvo} indicate a canonical average over a system of $N+1$ solvent molecules in a volume $V$, while $\De U_{o|i}(\br^{N+1})$ indicates the change in the interaction potential when converting solvent molecule $i$ into an $o$-solute.
This last average may be computed via the standard free energy perturbation (FEP) method, i.e., simulating a system of $N+1$ solvent molecules and averaging   
               $ \frac{1}{N+1}
                    \sum_{i=1}^{N+1}
                    \exp\left[
                    -\bt \De U_{o|i}(\br_t^{N+1})
                    \right]$
over the sampled configurations, $\br_t^{N+1}$.
One expects that, for sufficiently large systems limit, this ratio should depend only upon the density of solvent molecules, i.e., $\Yvo(\Nstvwz+1,\Vstvz) \rightarrow \Yvo^{\infty}(\rhostvwz)$ where $\rhostvwz = \Nstvwz/\Vstvz$ is the density of the vapor phase for a system of pure solvent at liquid-vapor coexistence.
Accordingly, we estimated $\Yvo^{\infty}(\rhostvwz)$ by evaluating $\Yvo(N_{vw},V_v)$ based upon a simulation of $N_{vw} = 100$ solvent molecules in a volume $V_v = N_{vw}/\rhostvwz$.
We approximated $\Yvo(\Nstvwz+1,\Vstvz)$ based upon this estimate for $\Yvo^{\infty}(\rhostvwz)$.
We then determined $\qvo$ from Eqs.~\eqref{def-qvo-methods}--\eqref{def-Yvo}.
However, it is not necessary to explicitly determine $\muwnull$ because the factor $e^{-\bt \muwnull}$ cancels out of every observable that we consider in this work.

\subsubsection{Liquid phase}
\label{subsubsec-FreeEnergyMethods-Liquid}
We followed the same approach to estimate the effective partition function, $\qlo$, for adding one $o$-solute to the liquid phase:
\begin{equation}
\label{def-qlo-methods}
\qlo    
    \equiv 
            \frac{1}{\Nstlwz} \frac{\Do(\Nstlwz,\Pnull)}{\Dnull(\Nstlwz,\Pnull)}
    =
        \left(\frac{\Nstlwz+1}{\Nstlwz}\right) 
            \left( \frac{\Law}{\Lao}\right)^3 
                \Ylo(\Nstlwz+1,\Pnull)    
                 e^{-\bt \muwnull}
            ,
\end{equation}
where $\Nstlwz$ is the number of solvent molecules in the liquid region in the bulk ensemble for pure solvent and $\Pnull$ is the vapor pressure of pure solvent.
$\Ylo(N+1,\Pnull)$ is defined by the corresponding ratio of configuration and volume integrals for the constant NPT ensemble, $\hZ_{po}(N,\Pnull)/\hZ_{p\phi}(N+1,\Pnull)$:
\begin{eqnarray}
\label{def-Ylo}
\Ylo(N+1,\Pnull)   
    & \equiv &  
        \frac{\hZ_{po}(N,\Pnull)}{\hZ_{p\phi}(N+1,\Pnull)}
    =
      \llangle 
        \frac{1}{N+1}
        \sum_{i=1}^{N+1}
            \exp\left[
                    -\bt \De U_{o|i}(\br^{N+1})
                    \right]
      \rrangle_{N+1,\Pnull,\phi}       .
\end{eqnarray}
Here the angular brackets denote an average in the isothermal-isobaric ensemble for a system with $N+1$ solvent molecules at a constant external pressure $\Pext = \Pnull$.
We expect that for sufficiently large systems $\Ylo$ should be independent of $N$, i.e., $\Ylo(N+1,\Pnull) \rightarrow \Ylo^\infty(\Pnull)$.
Accordingly, we estimated $\Ylo^\infty(\Pnull)$ by computing $\Ylo(\Nstlwz+2,\Pnull)$ for a system of $1994 \approx \Nstlwz+2$ solvent molecules.
We then approximated $\Ylo(\Nstlwz+1,\Pnull) \approx \Ylo^\infty(\Pnull)$ and determined $\qlo$ from Eq.~\eqref{def-qlo-methods}.
Given $\qvo$ and $\qlo$, we determined $\rKo$ from Eq.~\eqref{def-rKo-maintext}.
We then determined $\alpo$ and $\qbo$ from Eqs.~\eqref{eq-alpo-approx} and \eqref{def-alpo}.

The effective partition function for introducing two solutes into the liquid phase is 
\begin{eqnarray}
Z_{l\oop}	
    & \equiv & 	
        \Coop \qlo\inv \qlop \inv 
            \frac{\De_{\oop}(\Nstlwz,\Pnull)}{\Dnull(\Nstlwz,\Pnull)}
    =
        \Coop \qlo\inv \qlop \inv 
            \frac{\De_{\oop}(\Nstlwz,\Pnull)}{\Dnull(\Nstlwz+2,\Pnull)}
            e^{-2\bt\muwnull}
        ,
\end{eqnarray}
where we have introduced $1 = \Dnull(\Nstlwz+2,\Pnull)/\Dnull(\Nstlwz+2,\Pnull)$ and recognized $\Dnull(\Nstlwz+2,\Pnull) / \Dnull(\Nstlwz,\Pnull) = e^{-2 \bt \muwnull}$.
We express the remaining ratio of partition functions in analogy to Eq.~\eqref{eq-Qo-ratio-methods}:
\begin{eqnarray}
\Coop \frac{\De_{\oop}(\Nstlwz,\Pnull)}{\Dnull(\Nstlwz+2,\Pnull)}
    =
         \left(\frac{(\Nstlwz+2)!}{\Nstlwz!}\right) 
            \left( \frac{\Law^2}{\Lao\La_{o'}}\right)^3 
                \Yloop(\Nstlwz+2,\Pnull)    ,
\end{eqnarray}
where 
\begin{eqnarray}
\Yloop(N+2,\Pnull)    
    \equiv
        \frac{\hZ_{poo'}(N,\Pnull)}{\hZ_{p\phi}(N+2,\Pnull)}
    =
      \llangle 
        \frac{N!}{(N+2)!}
        \sum_{i=1}^{N+2}
        \sum_{j (\neq i)}^{N+2}
            \exp\left[
                    -\bt \De U_{oo'|ij}(\br^{N+2})
                    \right]
      \rrangle_{N+2,\Pnull,\phi}       
      \hspace{-1.3cm}
      . 
      \end{eqnarray}
Here $\De U_{oo'|ij}(\br^{N+2})$ is the change in the interaction potential when solvent molecules $i$ and $j$ are  converted to $o$ and $o'$-solutes, respectively. 
The two-solute effective partition function may then be expressed
\begin{eqnarray}
Z_{l\oop} 
    = 
        \left( \Nstlwz \right)^2
        \left( \frac{\Nstlwz+2}{\Nstlwz+1} \right)
        \frac{\Yloop(\Nstlwz+2,\Pnull)}
                {\Ylo(\Nstlwz+1,\Pnull)Y_{lo'}(\Nstlwz+1,\Pnull)}    
        .
\end{eqnarray}
We determined $\Yloop(\Nstlwz+2,\Pnull)$ via FEP based upon a simulation of $1994 \approx \Nstlwz+2$ solvent molecules. 
As noted above, we approximated $\Ylo(\Nstlwz+1,\Pnull)$ by $\Ylo(\Nstlwz+2,\Pnull)$ in order to eliminate 1-solute contributions from the 2-solute exchanges in $\Yloop(\Nstlwz+2,\Pnull)$.
We determined the interaction parameter for the liquid phase according to
$
\oveloop = - \left[ Z_{l\oop} - (\Nstlwz)^2 \right]  / \Nstlwz.
$
We then determined the bulk interaction parameter, $\oveboop$, from Eq.~\eqref{eq-oveboop-approx}.

\subsubsection{Inhomogeneous total system}
\label{subsubsec-FreeEnergyMethods-Total}
We similarly treated the effective partition function, $\qto$, for adding one $o$-solute to the inhomogeneous total system: 
\begin{equation}
\label{def-qto-methods}
\qto    
    \equiv 
            \frac{1}{\Nw} \frac{\Qto(\Nw,V,\si)}{\Qtnull(\Nw,V,\si)} 
    =
        \frac{\Nw+1}{\Nw} 
            \left( \frac{\Law}{\Lao}\right)^3 
                Y_{to}(\Nw+1,V,\si)
                 e^{-\bt \muwnull} 
            .
\end{equation}
Here we introduced the corresponding ratio of configuration integrals for the total system:
\begin{eqnarray}
\label{eq-Yto}
\Yto(N+1,V,\si)   
    & \equiv &  
        \frac{\hZ_{to}(N,V,\si)}{\hZ_{t\phi}(N+1,V,\si)}
    =
      \llangle 
        \frac{1}{N+1}
        \sum_{i=1}^{N+1}
            \exp\left[
                    -\bt \De U_{o|i}(\br^{N+1})
                    \right]
      \rrangle_{N+1,V,\si,\phi}       
     \hspace{-1.0cm} 
     ,
\end{eqnarray}
where the angular brackets denote an average for a system of $N+1$ solvent molecules in an inhomogeneous total system with volume $V$ and surface area $\si$.
Given $\qto$, $\qbo$, and $\qlo$, we then determined $\rto$ and $\ko$ from Eqs.~\eqref{def-rto} and \eqref{def-ko}.

The effective partition function for introducing two solutes into the inhomogeneous total system is 
\begin{eqnarray}
Z_{t\oop}	
    & \equiv & 	
        \Coop \qto\inv \qtop \inv 
            \frac{\Qtoop(\Nw,V,\si)}{\Qtnull(\Nw,V,\si)}
            \\
   & = &  
        \Nw^2
        \left(\frac{\Nw+2}{\Nw+1}\right) 
        \frac{\Ytoop(\Nw+2,V,\si)}{\Yto(\Nw+1,V,\si)Y_{to'}(\Nw+1,V,\si)}    
        ,
\end{eqnarray}
where 
\begin{eqnarray}   
\Ytoop(N+2,V,\si)    
    \equiv
        \frac{\hZ_{t\oop}(N,V,\si)}{\hZ_{t\phi}(N+2,V,\si)}
    =
      \llangle 
        \frac{N!}{(N+2)!}
        \sum_{i=1}^{N+2}
        \sum_{j (\neq i)}^{N+2}
            \exp\left[
                    -\bt \De U_{oo'|ij}(\br^{N+2})
                    \right]      
      \rrangle_{N+2,V,\si,\phi}       
      ,
\end{eqnarray}
We determined the interaction parameter, $\ovetoop$, for the total system according to Eq.~\eqref{def-ovetoop}.
Finally, we determined  $\hoop$ from Eqs.~\eqref{def-oveioop} and \eqref{def-hoop}.

\subsubsection{Eliminating 1-solute contributions from 2-solute calculations}
\label{subsubsec-FreeEnergyMethods-OneSoluteContributions}
The thermodynamic interaction parameters $\oveloop$ and $\ovetoop$ are both calculated from ratios of the form $Y\inv_o(N+1) Y\inv_{o'}(N+1) Y_{\oop}(N+2)$.
Here $Y_{\oop}(N+2)$ is calculated for a system of $N+2$ solvent molecules by converting two molecules into solutes, while $Y_o(N+1)$ and $Y_{o'}(N+1)$ are calculated for systems with $N+1$ solvent molecules by converting one molecule into a solute.
Our calculations of $\oveloop$ and $\ovetoop$ converged much more rapidly when $Y_o$, $Y_{o'}$, and $Y_{\oop}$ were all calculated from the same set of configurations. 
Consequently, we estimated $Y_o(N+1)$ based upon simulations with $N+2$ particles.

This estimate is straightforward for the liquid phase considered in subsection~\ref{subsubsec-FreeEnergyMethods-Liquid}.
For sufficiently large systems, we expect that $\Ylo(N,\Pnull)$, should be independent of $N$ because at constant external pressure the system will simply expand to accommodate the additional particle. 
Thus, $\Ylo(\Nstlwz+1,\Pnull) \rightarrow \Ylo^\infty(\Pnull)$.
Accordingly, we estimated $\Ylo^\infty(\Pnull)$ by calculating $\Ylo(\Nstlwz+2,\Pnull)$ for a system of $\Nstlwz+2$ molecules and then approximated $\Ylo(\Nstlwz+1,\Pnull)$ by this estimate for $\Ylo^\infty(\Pnull)$, i.e., $\Ylo(\Nstlwz+1,\Pnull)\approx \Ylo(\Nstlwz+2,\Pnull)$.

\newcommand{\Yloinf}{\Ylo^\infty}
\newcommand{\Yvoinf}{\Yvo^\infty}
\newcommand{\Yio}{Y_{io}}
\newcommand{\Yioinf}{\Yio^\infty}

However, this estimate is less straightforward for the inhomogeneous total system because adding an additional solvent molecule to the fixed volume will perturb the ratio of molecules in the liquid and vapor environments. 
We expect that the $\Nw+1$ Boltzmann factors, $\exp\left[ - \bt \De U_{o|i}(\br^{\Nw+1}) \right]$, that contribute to $\Yto(\Nw+1)$ in Eq.~\eqref{eq-Yto} sample qualitatively different environments in the liquid, vapor, and interfacial regions of the inhomogeneous total system. 
We assume that, on average, molecules that are in the liquid region and far from the interface will contribute $\Yloinf \equiv \Yloinf(\Pnull)$ to $\Yto$.
Similarly, we assume that, on average, molecules that are in the vapor region and far from the interface will contribute $\Yvoinf \equiv \Yvoinf(\rhostvwz)$ to $\Yto$.
Consequently, we adopted the following ansatz 
\begin{equation}
\label{eq-Yto-ansatz}
Y_{to}(\Nw+1) 
    = 
		\frac{\tNl}{\Nw+1} Y_{lo}^\infty 
	+ 
		\frac{\tNv}{\Nw+1} Y_{vo}^\infty 
	+ 
		\frac{\tNi}{\Nw+1} Y_{io}^\infty		,
\end{equation}   
where 
$\Yioinf$ is the contribution from molecules near the interface.
Here, we have partitioned the volume $V$ and the $\Nw+1$ solvent molecules in the slab system into 3 regions:
(1) a region of volume $\tVl$ containing $\tNl = \tVl \rhostlwz$ solvent molecules (on average) that behave like bulk liquid far from the interface; 
(2) a region of  volume $\tVv$ containing $\tNv = \tVv \rhostvwz$ solvent molecules (on average) that behave like bulk vapor far from the interface; 
and 
(3) a region of volume $\tVi = V - \tVl - \tVv$ containing $\tNi = \Nw + 1 -\tNl - \tNv$ solvent molecules that are influenced by the interface.
We emphasize that this partitioning does not correspond to the Gibbs equimolar partitioning, 
e.g., $\tVl \neq \Vstlz$.
Rather, it reflects the different environments that contribute to the average in Eq.~\eqref{eq-Yto}.

We are interested in evaluating $Y_{to}(\Nw+1)$ from a slab simulation of $\Nw+2$ solvent molecules in the volume, $V$.
We assume that the additional solvent molecule leaves the interfacial region unchanged, i.e., $\tVi$, and $\tNi$ do not change. 
Consequently, the volume of the vapor and liquid regions change by $\tVl \rightarrow \tVl + \de\tVl$ and $\tVv \rightarrow \tVv + \de\tVv$, respectively, with $\de\tVv = - \de\tVl$.
Moreover, the added solvent molecule partitions between the liquid and vapor regions according to 
\begin{equation}
1 = \de\tVl \, \rhostlwz + \de \tVv \, \rhostvwz = \de\tVl \, \Drhownull, 
\end{equation}
where $\Drhownull = \rhostlwz - \rhostvwz$.
Thus, we see that $\de\tVl = 1/\Drhownull = - \de\tVv$ and that the number of molecules in the liquid and vapor regions change by 
$
\de\tNl = 
        \rhostlwz / \Drhownull
$
and 
$
\de\tNv =
        - \rhostvwz / \Drhownull
$,
respectively.
Our ansatz for $\Yto$ implies that
\begin{equation}
\label{eq-tqto-Nw+2}
Y_{to}(\Nw+2)  
	= 
		\left( \frac{\tNl+\de\tNl}{\Nw+2} \right) \Ylo^\infty 
	+ 
		\left( \frac{\tNv+\de\tNv}{\Nw+2} \right) \Yvo^\infty 
	+ 
		\left( \frac{\tNi}{\Nw+2} \right) Y_{io}^\infty 		.
\end{equation}
By comparing with Eq.~\eqref{eq-Yto-ansatz}, we have
\begin{equation}
\label{eq-tqto-Nw+1}
Y_{to}(\Nw+1)  
	= 
		\left( \frac{\Nw+2}{\Nw+1} \right)	\Yto(\Nw+2)
	- 
		\frac{1}{\Nw+1}	
			\left[ 
			\de\tNl \, \Ylo^\infty + \de\tNv \, \Yvo^\infty
			\right]		.
\end{equation}
Thus, we determined $\Yto(\Nw+2)$ by evaluating the FEP expression in Eq.~\eqref{eq-Yto} with $\Nw+1 \rightarrow \Nw+2$. 
We employed Eq.~\eqref{eq-tqto-Nw+1} to determine $\Yto(\Nw+1)$.
We then calculated $\qto$ from Eq.~\eqref{def-qto-methods}.
